\documentclass[12pt]{article}
\pdfoutput=1
\usepackage{jheppub}
\usepackage[skip=2pt,font=footnotesize]{caption}
\usepackage[utf8]{inputenc}
\usepackage{verbatim}
\usepackage{amsmath}
\usepackage{amssymb}
\usepackage{amsthm}
\usepackage{slashed}
\usepackage{amsfonts}
\usepackage{makecell}
\usepackage{tabularx}
\usepackage{comment}
\usepackage[section]{placeins}
\usepackage{subfigure}
\usepackage{float}
\usepackage{color}
\usepackage{physics}
\usepackage{algpseudocode}
\usepackage{caption}
\usepackage{algorithm}
\usepackage{rotating}
\usepackage{multirow}

\newtheorem{theorem}{Theorem}

\begin{document}


\title{Quantum Power Flows: From Theory to Practice}

\author[a,b,c,d]{Junyu Liu$^{*,\$,}$}
\author[b,e]{Han Zheng$^{\$,}$}
\author[b,f]{Masanori Hanada}
\author[b,d]{Kanav Setia}
\author[g]{Dan Wu$^*$}

\affiliation[a]{Pritzker School of Molecular Engineering, The University of Chicago, Chicago, IL 60637, USA}
\affiliation[b]{qBraid Co., Harper Court 5235, Chicago, IL 60615, USA}
\affiliation[c]{Kadanoff Center for Theoretical Physics, The University of Chicago, Chicago, IL 60637, USA}
\affiliation[d]{Chicago Quantum Exchange, Chicago, IL 60637, USA}
\affiliation[e]{Department of Computer Science, The University of Chicago, IL, 60637}
\affiliation[f]{Department of Mathematics, University of Surrey, Guildford, Surrey GU2 7XH, United Kingdom}
\affiliation[g]{Lab for Information \& Decision Systems (LIDS), Massachusetts Institute of Technology, Cambridge, MA 02139, USA}

\hbox{$*$: Corresponding Authors.}
\hbox{$\$$: Equal contributions.}

\abstract{Climate change is becoming one of the greatest challenges to the sustainable development of modern society. Renewable energies with low density greatly complicate the online optimization and control processes, where modern advanced computational technologies, specifically quantum computing, have significant potential to help. In this paper, we discuss applications of quantum computing algorithms toward state-of-the-art smart grid problems. We suggest potential, exponential quantum speedup by the use of the Harrow-Hassidim-Lloyd (HHL) algorithms for sparse matrix inversions in power-flow problems. However, practical implementations of the algorithm are limited by the noise of quantum circuits, the hardness of realizations of quantum random access memories (QRAM), and the depth of the required quantum circuits. We benchmark the hardware and software requirements from the state-of-the-art power-flow algorithms, including QRAM requirements from hybrid phonon-transmon systems, and explicit gate counting used in HHL for explicit realizations. We also develop near-term algorithms of power flow by variational quantum circuits and implement real experiments for 6 qubits with a truncated version of power flows. }

\maketitle

\section{Power flow in renewable-rich smart grids}

\subsection{Motivation}
Climate change is becoming one of the greatest challenges to the sustainable development of modern society. A major contributor is believed to be the greenhouse gases emitted to the atmosphere by human activities \cite{manabe:1975effects,manabe:1991transient,manabe:1992transient}. In order to avoid irreversible climate change, greenhouse gas emissions shall be reduced to a certain level in the next few decades \cite{masson:2018global}, which requires comprehensive decarbonization of existing energy consumption sectors. 

Currently, the three leading sectors in greenhouse gas emissions are transportation ($27 \%$), electricity ($25 \%$), and industry ($24 \%$) \cite{epa}, occupying over three-quarters of the total greenhouse gas emission in 2020. The transportation and industry sectors are undergoing re-electrification upgrades to reduce greenhouse gas emissions by replacing fossil fuel-based units with electric-driven ones. A typical example is the popularization of electric vehicles. These upgrades, however, depend on the success of decarbonization in electric power systems to reduce the overall societal carbon emission.

An ongoing reform to realize power system decarbonization relies on large-scale implementations of renewable energy technologies because renewable generation emits very few greenhouse gases, if not none. However, high penetration of renewable energy into today's power grids can introduce various new engineering challenges, which, if not treated appropriately, can drastically degrade the quality of electricity service and incurs blackouts \cite{nerc:2017}. 

One challenge arises from the high intermittency and unpredictability of renewable energy. It can create much more unexpected operational scenarios that substantially exacerbate the computational burden for stability, reliability, and efficiency verification. Another challenge originates from the low density of renewable energy. To serve the future heavy load\footnote{As re-electrification unfolds, transportation and industry sectors will exert extra electric load demand to the existing power grids, creating extremely stressful demand in the near future. For example, by charging household electric vehicles in an unorganized manner, one can easily collapse the distribution system.}, a huge number of distributed inverter-based resources (IBR) have to be installed and coordinated in real-time, which greatly enlarges the system size and complicates the online optimization and control processes. The third challenge appeared recently when multi-component failures occurred more frequently in those renewable-based IBRs and finally evolved into blackouts \cite{eso:2020}. The traditional $N-1$ contingency analysis may no longer be able to secure the future renewable-rich smart grids. Instead, the $N-k$ contingency analysis should be carried out, which can easily create combinatorial contingencies. These new challenges will quickly drain the classical computation resources and, hence, call for a more powerful and efficient computational solution.

To address these computational challenges, we explore the possibility of using quantum algorithms to accelerate solving large-scale smart grid problems, from the theory to the real quantum hardware. Specifically, in this paper, we will focus on how to implement various theoretical or near-term quantum linear algorithms to solve a core problem in power engineering, \textit{the power-flow problem}. 
\subsection{Background of power-flow problem}
The power-flow problem (also known as the load-flow problem) comprises a set of nonlinear power-balancing equations derived from the energy conservation law at every node in the system. Solving a power-flow problem is to seek a feasible node voltage profile that satisfies all the power-balancing equations. It underpins almost all the further power system analysis and evaluations, including quasi-steady state analysis, state estimation, stability analysis and control, optimal dispatch and scheduling, unit commitment and planning, $N-1$ contingency analysis, reliability evaluation, etc. 
Early attempts to solve the power-flow problem by using classical digital computers started in the 1950s, when the Gauss-Seidel-like iterations were adopted \cite{ward:1956digital,mcgillis:1957nodal,jordan:1957rapidly}. However, this type of method requires restrictive conditions on the Jacobian matrix to guarantee convergence. In the 1960s, Newton's method was introduced to improve the convergence \cite{james:1961,tinney:1967power}. It then became one of the most popular methods of solving power-flow problems. In order to better improve computational efficiency, the fast decoupled load flow method was proposed in 1978 \cite{iwamoto:1978fast}. It is a quasi-Newton method specially tailored for solving power-flow problems under light and moderate loading conditions. In the recent decade, the holomorphic embedding method was developed and gained great popularity in the power community for solving power-flow and related problems \cite{trias:2012holomorphic, chiang:2017novel,wu:2019holomorphic,wang:2020theoretical}. Although it was claimed to be a non-iterative method, constructing the power-flow holomorphic embedding function is still an iterative process that requires repeatedly solving linear systems of equations. It transfers the requirement of the iterative process from the numerical solution evaluation to the holomorphic function construction. 

Therefore, almost all the existing power-flow solvers require repeatedly solving linear systems of equations. As the power-flow problem size increases substantially, this computation module can easily become one of the main bottlenecks for many stability, reliability, and efficiency evaluations in the future renewable-rich smart grid. 

Recently, a few research papers proposed the use of quantum algorithms and quantum hardware to accelerate various numerical evaluations of smart grids. In \cite{eskandarpour:2020quantum} the authors studied the theoretical possibility of using a quantum algorithm to solve $N-k$ power system contingency analysis. Then, \cite{feng:2021quantum} proposed the quantum power-flow model in the fast-decoupled framework. Meanwhile, \cite{zhou:2021quantum} proposed to use quantum algorithms in the electromagnetic transient program. However, these results were only verified on the simulated quantum hardware. To testify the existing quantum algorithms for power-system studies on real quantum hardware, \cite{saevarsson:2022quantum} studied the 3-node example in the lossless condition with explicitly known eigenvalues, while \cite{eskandarpour:2021experimental} studied the 9-node example in the DC power-flow model. 

All the above-mentioned results were based on the  Harrow-Hassidim-Lloyd (HHL) quantum linear solver \cite{harrow:2009quantum} which may be difficult for large-scale applications due to the hardness of practical implications. Moreover, the speed of real quantum hardware not only depends on algorithmic oracles but also heavily relies on the quantum memory uploading and downloading procedures, which have not been investigated extensively in the above literature. In this paper, improved analyses regarding these issues including the requirements of quantum random access memory on real quantum hardware will be presented. We will also discuss how to accelerate solving the power-flow problem by using an alternative noisy near-term quantum linear algorithm.
\subsection{Mathematical modeling of power-flow problem}
The standard power-flow problem can be represented in either the polar coordinate system or the Cartesian coordinate system. Throughout this paper, we formulate the power-flow problem in the Cartesian coordinate system. Specifically, consider a power grid with $N_{\textrm{bus}}$ many buses\footnote{We call a node in the power grid a \textit{bus}.}. Without loss of generality, let the $1^{\textrm{st}}$ bus be the slack bus at which the voltage magnitude and the voltage angle are specified. Let the $2^{\textrm{nd}}$ to the $N_{\textrm{gen}}$-th bus be the PV bus at which the active power injection and the voltage magnitude are provided. Let the rest of the buses be the PQ bus at which both the active and the reactive power injections are given. Then, we have the following equations:
\begin{align}
k = 1:&\left\{ \begin{array}{l}
\mathbf{U}^T\mathbf{M}_{\textrm{v},1}\mathbf{U} - V_{k,\textrm{m}}^2 = 0\\
V_{\rm{q},k} = 0
\end{array} \right.\label{eq:PF_slack}\\
2 \le k \le N_{\textrm{gen}}:&\left\{ \begin{array}{l}
\mathbf{U}^T\mathbf{M}_{\textrm{p},k}\mathbf{U} - P_{\textrm{gen},k} = 0\\
\mathbf{U}^T\mathbf{M}_{\textrm{v},k}\mathbf{U} - V_{k,\textrm{m}}^2 = 0
\end{array} \right.\label{eq:PF_PV}\\
N_{\textrm{gen}} + 1 \le k \le N_{\textrm{bus}}:&\left\{ \begin{array}{l}
\mathbf{U}^T\mathbf{M}_{\textrm{p},k}\mathbf{U} + P_{\textrm{load},k} = 0\\
\mathbf{U}^T\mathbf{M}_{\textrm{q},k}\mathbf{U} + Q_{{\textrm{load}},k} = 0
\end{array} \right.\label{eq:PF_PQ}
\end{align}
where $\mathbf{U} \in \mathbb{R}^{2N_{\rm{bus}}}$ piles up all the real parts and the imaginary parts of bus complex voltages, e.g. at bus-$k$ the complex voltage is expressed as $\mathbf{U}_{2k-1}+\sqrt{-1}\mathbf{U}_{2k}$, that need to be solved. $V_{\rm{q},1}$ is the imaginary part of the complex voltage at bus-1, and hence the same as one of the components of $\mathbb{U}$ ($V_{\rm{q},1}=\mathbf{U}_2$ in the convention above). $\mathbf{M}_{\textrm{p},k},\mathbf{M}_{\textrm{q},k},\mathbf{M}_{\textrm{v},k} \in  \mathbb{R}^{2N_{\textrm{bus}}\times 2N_{\textrm{bus}}}$ ($k=1,\cdots,N_{\textrm{bus}}$) are given sparse constant matrices. 
$V_{k,\textrm{m}}$ ($k=1,\cdots,N_{\textrm{gen}}$), $P_{\textrm{gen},k}$ ($k=2,\cdots,N_{\textrm{gen}}$), $P_{\textrm{load},k}$ and $Q_{\textrm{load},k}$ ($k=N_{\textrm{gen}}+1,\cdots,N_{\textrm{bus}}$) are given constants. 
Thus, we have $2 N_{\text{bus}}$ equations with $2 N_{\text{bus}}$ variables in total\footnote{Usually we substitute the second equation of \eqref{eq:PF_slack} to all the other equations to eliminate one variable and one equation.}. 

For simplicity, we will denote the set of $2N_{\textbf{bus}}$ equations \eqref{eq:PF_slack}, \eqref{eq:PF_PV}, and \eqref{eq:PF_PQ} as $\mathbf{F}(\mathbf{U})=0$. Classically, we could solve the above equations by following Newton's method:

\begin{center}
\begin{algorithmic}
\State Algorithm: \textbf{Newton-Raphson}$(\mathbf{F}, \mathbf{U}_0, K_{\text{max}}, \epsilon_0)$
\\
\State Given $\mathbf{U}_0$ \Comment{Provide an initial guess}
\State Given $K_{\text{max}}$ \Comment{Define the maximum iteration number}
\State Given $k=0$ \Comment{Initialize the iteration index}
\State Given $\epsilon_0 >0$ \Comment{Accuracy tolerance}
\While{$k\le K_{\text{max}}$ or $|\mathbf{F}_k| \ge \epsilon_0$}
\State $\mathbf{F}_k\gets \mathbf{F}(\mathbf{U}_k)$ \Comment{Compute the mismatch}
\State $\mathbf{J}_k \gets \mathbf{J}(\mathbf{U}_k)= \frac{\partial\mathbf{F}(\mathbf{U})}{\partial \mathbf{U}} |_{\mathbf{U}=\mathbf{U}_k}$ \Comment{Update the Jacobian matrix}
\State $d\mathbf{U}_k \gets -\mathbf{J}_k^{-1} \mathbf{F}_k$ \Comment{Compute the increment of $\mathbf{U}$}
\State $\mathbf{U}_{k+1} \gets \mathbf{U}_k  + d\mathbf{U}_k$ \Comment{Compute new $\mathbf{U}$}
\State $k \gets k+1$ \Comment{Update the index $k$}
\EndWhile
\State Output $\mathbf{U}_k$
\end{algorithmic}
\end{center}

Schematically, the Jacobian can be written as
\begin{align}
\mathbf{J}(\mathbf{U}) = 
\frac{\partial \mathbf{F}(\mathbf{U})}{\partial \mathbf{U}} =
\left[ {\begin{array}{*{20}{c}}
\mathbf{M}_{\textrm{v},i} \mathbf{U},
 \cdots, 
\mathbf{M}_{\textrm{q},k} \mathbf{U},
 \cdots, 
\mathbf{M}_{\textrm{p},j} \mathbf{U},
 \cdots 
\end{array}} \right]^T \label{eq:PF_Jacobian}
\end{align}
The Jacobian $\mathbf{J}$ is a $2N_{\text{bus}}\times 2N_{\text{bus}}$ matrix. The bottleneck is the matrix inversion, and we have:
\begin{theorem}
The algorithm $\operatorname{Newton}$-$\operatorname{Raphson}(\mathbf{F}, \mathbf{U}_0, K_{\rm{max}})$ has the complexity,
\begin{align}
\mathcal{O}\left( {{K}_{\max }}{{N}_{\rm{bus}}}s\kappa \log \frac{1}{{{\text{ }\!\!\epsilon\!\!\text{ }}_{\rm{inverse}}}} \right)\
\end{align}
where $s$, $\kappa$, and $\epsilon_{\rm{inverse}}$ are the maximal sparsity, the maximal condition number, and the minimal inversion error during all iterations.\footnote{
The inversion error is defined so that the difference between the true solution $\vec{x}_{\rm true}$ and the approximate solution $\vec{x}_{\rm approx.}$ satisfies
$|\vec{x}_{\rm true}-\vec{x}_{\rm approx.}|< \epsilon_{\rm inverse}$. 
} 
\end{theorem}
The complexity estimates come from the classical algorithms for solving sparse matrix inversions. In this paper, we propose quantum algorithms that could accelerate this algorithm. 

\subsection{Sparsity and condition number of power-flow Jacobian matrix}
The sparsity and the condition number of the power-flow Jacobian matrix play an important role in the complexity analysis of solving power-flow problems for both classical algorithms (mentioned above) and quantum algorithms (shall be seen shortly below). 

The power-flow Jacobian matrix is very sparse due to the loose topology of bulk power grids. From engineering heuristics, the average nodal degree of a realistic power grid is usually below 3 with a maximum degree of around 15. Recall \eqref{eq:PF_Jacobian} that the $i$-th row of the Jacobian matrix is given by
\begin{equation}
    \mathbf{J}_{i} = \mathbf{U}^T \mathbf{M}_i
\end{equation}%
for some real symmetric matrix $\mathbf{M}_i$ and $i=1,2,\dots,2 N_{\textrm{bus}}$. Each $\mathbf{M}_i$ has only two nonzero columns and the corresponding transposed rows. In each nonzero column, the nonzero entries only appear in the position where the associated $l$-th bus directly connects to. So the total number of nonzero entries $r_i$ in the $i$-th row of the Jacobian matrix is
\begin{equation}
    r_i \le 2 d_l
\end{equation}
where $d_l$ is the degree of bus-$l$ for $l=1,2,\dots,N_{\textrm{bus}}$. 

On the other hand, the $i$-th voltage variable $\mathbf{U}_i$ only appears in those nodal power balancing equations whose buses directly connect to the $l$-th bus. Hence, the total number of nonzero entries $c_i$ in the $i$-th column of the Jacobian matrix is
\begin{equation}
    c_i \le 2 d_l
\end{equation}

\begin{figure}[tb!]
	\centerline{\includegraphics[width=0.7\columnwidth]{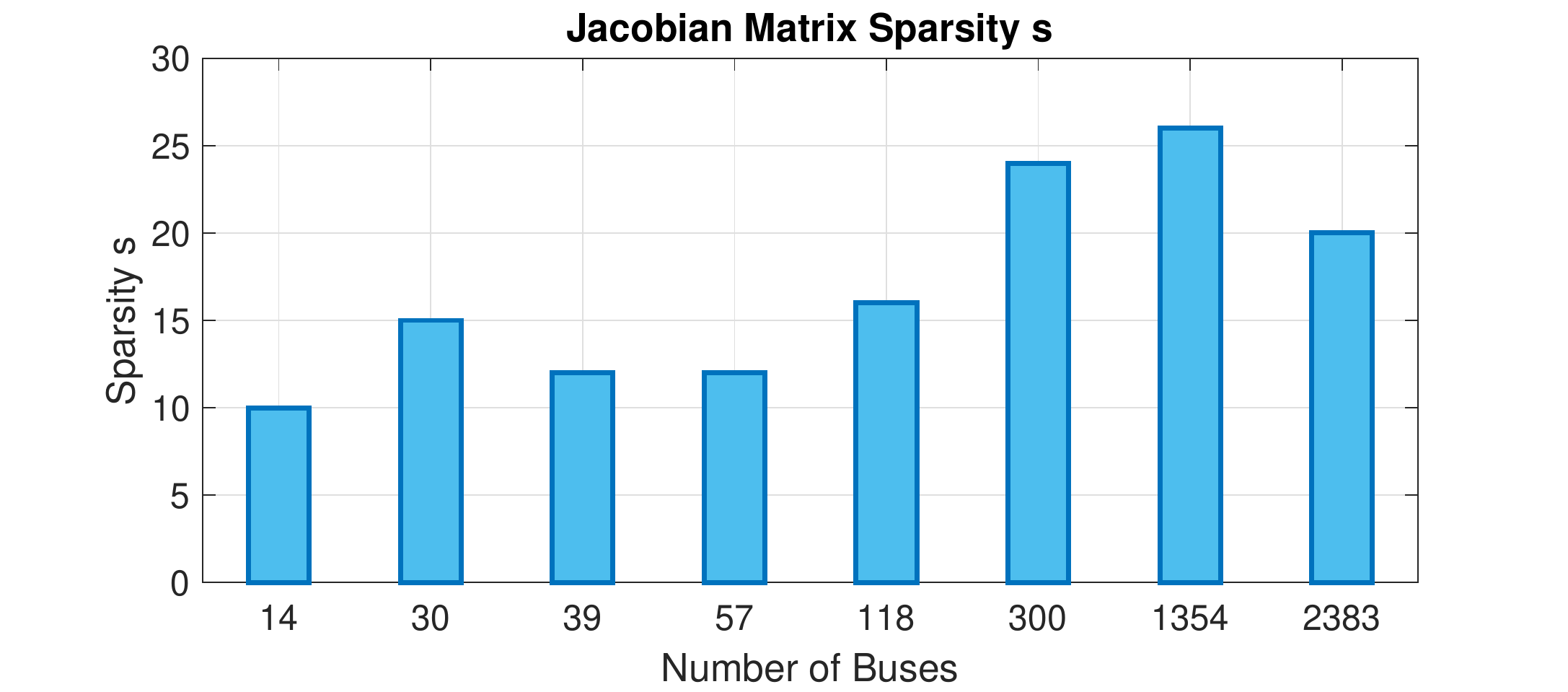}}
	\caption{Sparsity of Power-Flow Jacobian Matrix}\label{fig:sparsity}
\end{figure}

Therefore, the sparsity $s$ of $\mathbf{J}$ is
\begin{equation}
    s = \textrm{max} \{ r_i, ~c_i,~i=1,2,\dots,2 N_\textrm{bus} \} 
\end{equation}
which is usually around 20. In Figure~\ref{fig:sparsity} we demonstrate the sparsity values for some benchmark systems. Note that the sparsity does not depend significantly on the system size.

\begin{figure}[tb!]
	\centerline{\includegraphics[width=0.7\columnwidth]{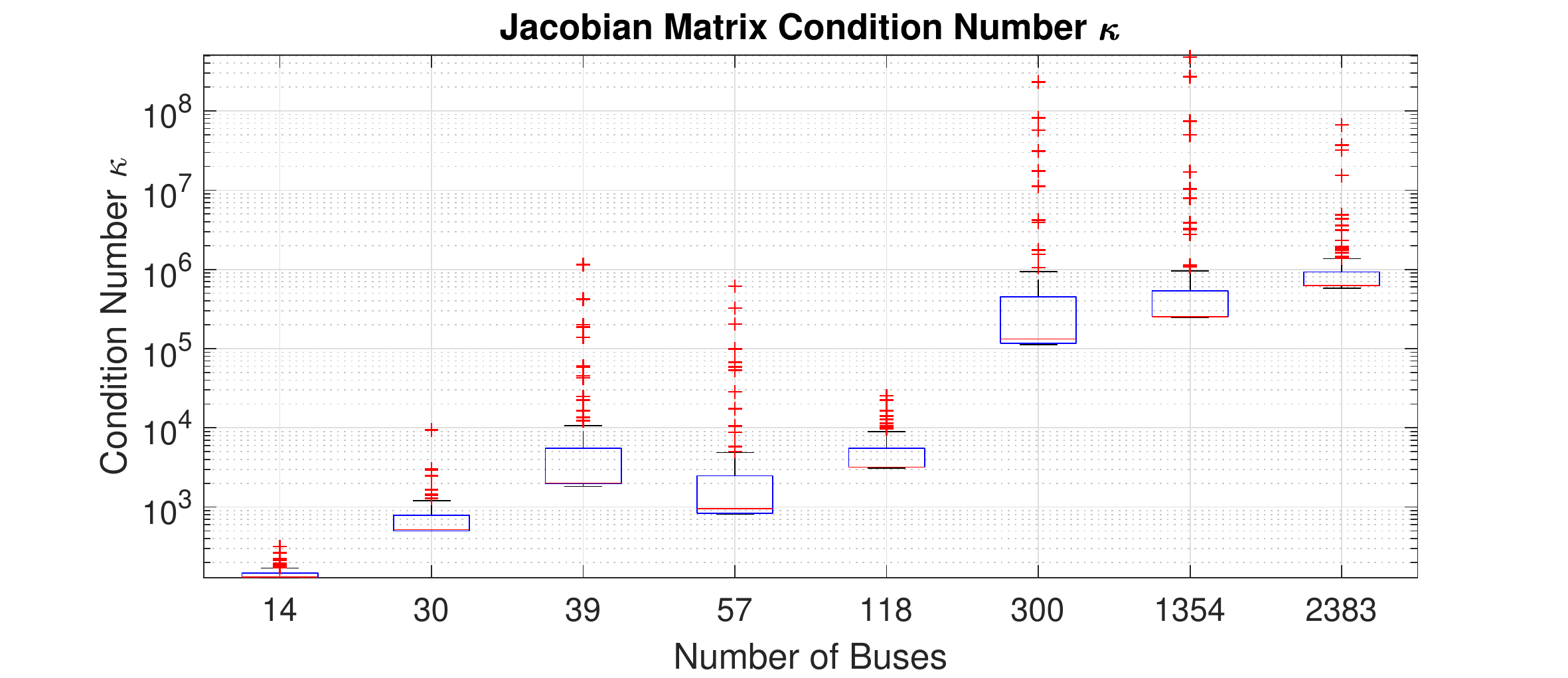}}
	\caption{Condition Number of Power-Flow Jacobian Matrix obtained from different steps of Newton-Raphson algorithm. The red line segment represents the medium value for each testing case. The bottom and top edges of the blue box indicate 25 and 75 percentiles, respectively. Red points are the outliers that are more than 50\% larger than the top of the blue box shown by the black line. 
  Typically, the condition number increases at first and then declines along the iteration sequence.}\label{fig:condition_number}
\end{figure}

The condition number of the Jacobian matrix is another important factor that can affect the complexity of solving power-flow problems. When approaching the loadability limit, the power-flow Jacobian matrix is near singular, yielding a very large condition number. However, under normal operating conditions, the condition number of a power grid can remain in a certain range. Figure~\ref{fig:condition_number} illustrates condition numbers of different scenarios for a few benchmark power systems. The condition numbers of most of the scenarios fall in the range of $10^6$ with a few outliers being up to $10^9$. 
Because the scaling of the condition number with respect to the system size $N_{\rm bus}$ can depend on the details of the system, we cannot find a rigorous estimation of the condition number for our power-flow problems. In the following, we treat the condition number $\kappa$ as an $O(N_{\rm bus}^0)$ quantity unless otherwise stated. 
\section{Uploading and downloading}
Before we discuss real-world applications, it is necessary for us to understand how to create interfaces between classical and quantum processors. In this section, we briefly explain quantum random access memory and classical shadow, which can be useful for uploading and downloading classical data.
\subsection{Uploading: quantum random access memory}
Quantum Random Access Memory (QRAM) \cite{giovannetti2008quantum} provides convenient interfaces from classical to quantum processors. More precisely, a QRAM is the physical realization of the data-looking oracle
\begin{align}
    \sum_{i=0}^{N-1}\alpha_i\ket{i}\ket{0}
    \to
    \sum_{i=0}^{N-1}\alpha_i\ket{i}\ket{x_i}
    \label{data-lookup-oracle}
\end{align}
with the data size $N$, the data $x_i$, quantum pointer $\ket{i}$, and arbitrary coefficients $\alpha_i$. For given data with the size $N$, QRAM will use $\mathcal{O}(\log N)$ switches with $\mathcal{O}(N)$ qubits to upload the data. QRAM is a useful quantum architecture to realize quantum oracle constructions required by quantum algorithms, where the Harrow-Hassidim-Lloyd (HHL) algorithm \cite{harrow:2009quantum} is a typical example. Quantum oracles could be specified in different forms. For instance, one could directly use QRAM to realize the data lookup oracle, and the linear-combinations-of-unitaries (LCU) decomposition \cite{hann2021practicality}. 

Realizing QRAMs, in reality, is a possible but challenging task. Although proposals with experiments are investigated, making large-scale, fault-tolerant QRAMs is still in development \cite{hann2021resilience}. In this work, we will assume that there are possible QRAMs or other data uploaders existing for our quantum algorithms for power flows. A detailed investigation of hardware is given in Sec.~\ref{qram_require}. 
\subsection{Downloading: classical shadows}\label{sec:download}
Shadow tomography, discussed recently in \cite{huang2020predicting}, is a useful method for reading sparse quantum states from a quantum processor. The foundation of the classical shadow is based on the following theorem:
\begin{theorem}[Informal]
One can compute $\operatorname{Tr}(\rho O_i)_{i=1}^M$ with $M$ operators $O_i$, by only accessing the state $\rho$ 
\begin{align}
\mathcal{O}\left( \log (M){{\max }_{i}}\frac{\left\| {{O}_{i}} \right\|_{\operatorname{shadow }}^{2}}{{{\epsilon }^{2}}} \right)
\end{align}
times, where $\left\| {{O}_{i}} \right\|_{\operatorname{shadow }}^{2}$ is a version of the operator norm, and $\epsilon$ is the demanded error. We require $M$ to be not exponentially large against the number of qubits used in the state $\rho$. 
\end{theorem}
Considering the situation where we choose $O_i$ to be Pauli operators, we could obtain coefficients on a computational basis. 
By using an appropriate scheme (``Clifford measurement" discussed in \cite{huang2020predicting} ) we can make ${{\max }_{i}}\left\| O_i \right\|_{\operatorname{shadow }}^{2} $ of order one and independent of the system size.  

Suppose a given quantum state is sparse, and we do not have to take the error $\epsilon$ to be exponentially small to determine the state with good precision. 
The results of $\mathcal{O} (\log M/\epsilon^{-2})$ measurements are downloaded, and the evaluation of $M$ values is processed on classical devices. $\log M$ is at most the number of qubits. The tasks on classical devices can easily be parallelized. Hence shadow tomography is a useful way to download sparse quantum states from quantum processors to classical devices. 

If the state is not sparse, the downloading process can become exponential, and it is hard to maintain the quantum computational advantage. In the future, a better method for downloading would resolve this issue. In Sec.~\ref{sec:VQPF}, we present an algorithm that does not have such a potential issue associated with the downloading process. 

Note also that this downloading procedure is trivially parallelizable in the sense that, in the case of QPF-HHL algorithm discussed in Sec.~\ref{sec:QPF-HHL}, we can run the HHL algorithm separately on multiple quantum devices and use different random unitaries on different devices to take different shadows. There are no communications needed. If we use $n$ quantum devices, we can finish the task $n$ times faster. Often, such a parallelization is difficult for the solver on a classical device.

\section{Quantum algorithms}
\subsection{The HHL-based algorithm (QPF-HHL)}\label{sec:QPF-HHL}
Here, we describe quantum algorithms with a guaranteed computational advantage against the classical algorithm. The HHL algorithm \cite{harrow:2009quantum} provides the exponential guaranteed speedup against classical algorithms in solving linear equations $A\vec{x}=\vec{b}$ and finding $\vec{x}=A^{-1}\vec{b}$, where an $N\times N$ sparse Hermitian matrix $A$ and an $N$-component vector $\vec{b}$ are given. The HHL algorithm could be directly applied here when we are evaluating the Jacobian inverse. We would denote the HHL algorithm as \textbf{HHL}$(A,\ket{b},\epsilon_{\text{inverse}})$, where the algorithm will output a solution $\ket{x}$ where $\ket{x}=A^{-1}\ket{b}$. Here, $\ket{x}$ and $\ket{b}$ are defined by
\begin{align}
\ket{x}=\frac{\sum_{i=1}^Nx_i\ket{i}}{\sqrt{\sum_{i=1}^N|x_i|^2}}, 
\quad
\ket{b}=\frac{\sum_{i=1}^Nb_i\ket{i}}{\sqrt{\sum_{i=1}^N|b_i|^2}}.  
\end{align}
Importantly, the output is a quantum state. One could use classical shadow techniques \cite{huang2020predicting} to reconstruct the state using Pauli measurements, as discussed before. 

To understand the HHL algorithm, let us introduce an inefficient classical algorithm that can be made efficient on quantum devices! Let $\vec{u}_i$ and $\lambda_i$ ($i=1,2,\cdots,N$) be eigenvector and eigenvalue of $A$, satisfying $A\vec{u}_i=\lambda_i\vec{u}_i$. Once we determine all eigenvectors, we can write $\vec{b}$ as a linear combination of them as $\vec{b}=\sum_{i=1}^N\beta_i\vec{u}_i$. Because $A^{-1}\vec{u}_i=\frac{1}{\lambda_i}\vec{u}_i$, we obtain 
\begin{align}
\vec{x}=A^{-1}\vec{b}=\sum_{i=1}^N \frac{\beta_i}{\lambda_i}\vec{u}_i. 
\end{align}
This algorithm is very simple, but it is costly to determine all $\vec{u}_i$ and $\lambda_i$. Very interestingly, however, on quantum devices, this procedure can be done without explicitly constructing $\vec{u}_i$ and $\lambda_i$. This is the key idea of the HHL algorithm. 

The outline of the HHL algorithm is as follows. We assume $\ket{b}$ is given. Although we do not know the eigenstates $\ket{u_i}$, we know that $\ket{b}$ can be written as $\ket{b}=\frac{1}{\sqrt{\sum|\beta_i|^2}}\sum_{i=1}^N\beta_i\ket{u_i}$, with some unknown coefficients $\beta_i$. Although we do not know the eigenvalues $\lambda_i\in\mathbb{R}$, we can build a state $\frac{1}{\sqrt{\sum|\beta_i|^2}}\sum_{i=1}^N\beta_i\ket{u_i}\ket{\lambda_i}$ via the Quantum Phase Estimation (QPE) algorithm, where $\ket{\lambda_i}$ is the binary representation of $\lambda_i$.  
We add an ancilla qubit and obtain $\frac{1}{\sqrt{\sum|\beta_i|^2}}\sum_{i=1}^N\beta_i\ket{u_i}\ket{\lambda_i}\ket{0}$.
It is possible to construct an operator that generates 
\begin{align}
\ket{\lambda}\ket{0}
\to
\ket{\lambda}\left(
\sqrt{1-\frac{C^2}{\lambda^2}}\ket{0}
+
\frac{C}{\lambda}\ket{1}
\right). 
\label{eigenvalue-inversion}
\end{align}
Here $C$ is taken as large as possible while satisfying $|C/\lambda|<1$ for $\lambda=\lambda_1,\cdots,\lambda_N$. 
By acting this operator to $\frac{1}{\sqrt{\sum|\beta_i|^2}}\sum_{i=1}^N\beta_i\ket{u_i}\ket{\lambda_i}\ket{0}$, we obtain
\begin{align} \label{eq:hhl-inversion}
\frac{1}{\sqrt{\sum|\beta_i|^2}}\sum_{i=1}^N\beta_i\ket{u_i}
\ket{\lambda_i}\left(
\sqrt{1-\frac{C^2}{\lambda_i^2}}\ket{0}
+
\frac{C}{\lambda_i}\ket{1}
\right). 
\end{align}
We perform the inverse QPE to uncompute $\ket{\lambda_i}$ so that we obtain
\begin{align}
\frac{1}{\sqrt{\sum|\beta_i|^2}}\sum_{i=1}^N\beta_i\ket{u_i}
\left(
\sqrt{1-\frac{C^2}{\lambda_i^2}}\ket{0}
+
\frac{C}{\lambda_i}\ket{1}
\right). 
\end{align}

Then we measure the ancilla qubit. If we do not observe $\ket{1}$, we go back to the beginning and construct this state again. When $\ket{1}$ is observed, then the state in our hands is 
\begin{align}
\left(
\sum_{i=1}^N\left|\frac{\beta_i}{\lambda_i}\right|^2
\right)^{-1/2}
\sum_{i=1}^N\frac{\beta_i}{\lambda_i}
\ket{u_i}. 
\end{align}
This is the state $\ket{x}$ that corresponds to $\vec{x}=A^{-1}\vec{b}$. The assumption that $A$ is sparse and Hermitian is needed for the QPE to be done efficiently.

If $A$ is not Hermitian, the we can use $\tilde{A}\equiv\left(
\begin{array}{cc}
0 & A\\
A^\dagger & 0
\end{array}
\right)$ and $\vec{\tilde{b}}\equiv \left(
\begin{array}{c}
\vec{b}\\
0
\end{array}
\right)$ instead of $A$ and $\vec{b}$. Then, the rest of the argument does not change. Note that Hermiticity is needed for the QPE.   

We can use $-\mathbf{J}_k$, $\mathbf{U}_k$, and $d\mathbf{U}_k$ as $A$, $\vec{b}$ and $\vec{x}$. An important feature of the HHL algorithm is that the solution can be determined only up to the overall normalization factor, i.e., we obtain $\left(\sum_{i=1}^N\left|\frac{\beta_i}{\lambda_i}\right|^2\right)^{-1/2}A^{-1}\vec{b}=\frac{\vec{x}}{|\vec{x}|}$ rather than $\vec{x}$. Therefore, we must determine the normalization factor $|\vec{x}|$. We will give a comment in Sec.~\ref{sec:normalization}. 

Thus, a natural modification of the algorithm is as follows.

\begin{center}
\begin{algorithmic}
\State Algorithm: \textbf{QPF-HHL}$(\mathbf{F}, \mathbf{U}_0, K_{\text{max}},\epsilon_{\text{inverse}},, \epsilon_0)$
\\
\State Given $\mathbf{U}_0$ \Comment{Provide an initial guess}
\State Given $K_{\text{max}}$ \Comment{Define the maximum iteration number}
\State Given $k=0$ \Comment{Initialize the iteration index}
\State Given $\epsilon_0 >0$ \Comment{Accuracy tolerance}
\While{$k\le K_{\text{max}}$ or $|\mathbf{F}_k| \ge \epsilon_0$}
\State $\mathbf{F}_k\gets \mathbf{F}(\mathbf{U}_k)$ \Comment{Compute the mismatch}
\State $\mathbf{J}_k \gets \mathbf{J}(\mathbf{U}_k)= \frac{\partial \mathbf{F}(\mathbf{U})}{\partial \mathbf{U}} |_{\mathbf{U}=\mathbf{U}_k}$ \Comment{Update the Jacobian matrix}
\State Upload the matrix $\mathbf{J}_k$ and the state $\mathbf{F}_k$ to the quantum computer using Quantum Random Access Memory (QRAM).
\State $\ket{d\mathbf{U}_k} \gets -\ket{\mathbf{J}_k^{-1} \mathbf{F}_k}$ \Comment{Compute the increment of $\mathbf{U}$ using the HHL algorithm \textbf{HHL}$(\mathbf{J}_k,\mathbf{F}_k,\epsilon_{\text{inverse}})$, and download $\ket{d\mathbf{U}_k} =$\textbf{HHL}$(\mathbf{J}_k,\mathbf{F}_k,\epsilon_{\text{inverse}})$ using the classical shadow.}
\State Determine $d\mathbf{U}_k$ from $\ket{d\mathbf{U}_k}$  
\State $\mathbf{U}_{k+1} \gets \mathbf{U}_k  + d\mathbf{U}_k$ \Comment{Compute new $\mathbf{U}$}
\State $k \gets k+1$ \Comment{Update the index $k$}
\EndWhile
\State Output $\mathbf{U}_k$
\end{algorithmic}
\end{center}
Note that we are focusing on the improvement of the bottleneck of the classical computation, which is the computation of $\mathbf{J}_k^{-1}\mathbf{F}_k$. 

$\mathbf{F}_k$ and $\mathbf{J}_k$ are calculated classically, and the computational cost is $\mathcal{O}(N_{\rm bus})$. However, the overall factor in front of $N_{\rm bus}$ is small and independent of the required precision. Furthermore, the evaluation of $\mathbf{F}_k$ and $\mathbf{J}_k$ can easily be parallelized on classical devices without involving many communications between the nodes. A similar statements holds when we determine $d\mathbf{U}_k$ from $\ket{d\mathbf{U}_k}$; see Sec.~\ref{sec:download}.

We have the following provable guarantees.
\begin{theorem}
The algorithm ${\operatorname{QPF-HHL}}(\mathbf{F},\mathbf{U}_0, K_{\max},\epsilon_{\operatorname{inverse}})$ has the complexity,
\begin{align}
\mathcal{O}\left( {{K}_{\max }}\times{\log{N}_{\rm{bus}}}\times s^2\kappa^2 \frac{1}{{{{ }\!\!\epsilon\!\!\text{ }}_{\rm{inverse}}^2}} \right)\
\label{HHL-cost}
\end{align}
where $s$, $\kappa$, and $\epsilon_{\rm{inverse}}$ are the maximal sparsity, the maximal condition number, and the minimal inversion error during all iterations. The scaling factor should also include the downloading and uploading costs of the algorithm.
\end{theorem}

We emphasize that the cost for the downloading $\ket{d\mathbf{U}_k}$ which was discussed in Sec.~\ref{sec:download} should be multiplied to \eqref{HHL-cost}. It might be challenging to maintain the same quantum advantage if we cannot download the states efficiently. Note also that there are other implementations of the HHL algorithm that have different dependencies on the condition number $\kappa$. If $\kappa$ increases with $N_{\rm bus}$, we should choose an implementation with a weaker $\kappa$-dependence. In such a case, although the exponential speedup is lost, the polynomial speedup could still be achieved. 

\subsection{The VQLS-based algorithm (QPF-VQLS)}
The HHL algorithm, although with provable guarantees of quantum advantage, is usually considered to be an algorithm that could only work in the fault-tolerant regime with quantum error correction codes. In the near term, variational alternatives of HHL have been proposed \cite{bravo:2019variational}, that are supposed to run in noisy near-term quantum computers. Here, in order to make a more realistic estimation of the possibility of the near-term impact of quantum algorithms on the power-flow problems, we adopt the Variational Quantum Linear Solver (VQLS) \cite{bravo:2019variational} into the power-flow problem setup.  

Variational quantum algorithms usually have the following logic~\cite{cerezo2021variational}. One could construct the so-called variational ans\"{a}tze as
\begin{align}
U(\vec{\theta})=\prod\nolimits_{\ell =1}^{L}{{{W}_{\ell }}{{U}_{\ell }}({{\theta }_{\ell }})}=\prod\nolimits_{\ell =1}^{L}{{{W}_{\ell }}\exp (i{{\theta }_{\ell }}{{X}_{\ell }})}~,
\end{align}
where $W_{\ell}$s are fixed unitary gates and $X_{\ell}$s are fixed Hermitian operators. A collection of variational angle $\theta_\ell$s denoted as a vector $\vec{\theta}$ is a classical object that parametrizes the quantum circuits. 
The spirit of variational ans\"{a}tze to find the approximate answer is similar to neural networks, which are considered universal approximators of arbitrary high-dimensional functions (Thus, variational quantum ans\"{a}tze are also called quantum neural networks; see \cite{farhi2018classification}). 

In the sparse matrix inversion problem, we could approximate the state $\ket{x}=\frac{\sum_ix_i\ket{i}}{\sqrt{\sum|x_i|^2}}$ that corresponds to the solution of $A \vec{x}=\vec{b}$ by $\ket{x}=U(\vec{\theta}) \ket{0}$, where $\ket{0}$ is a prepared initial state. We could make a loss function by taking the infidelity
\begin{align}
    \mathcal{L}_{\rm G}=1-|\bra{b}\ket{\psi}|^2~,
\end{align}
where
\begin{align}
    \ket{\psi}
    =
    \frac{AU(\vec{\theta})\ket{0}} {|AU(\vec{\theta})\ket{0}|}
\end{align}
is the normalized state proportional to $AU(\vec{\theta})\ket{0}$,  
and perform some classical optimization algorithms to find the optimal $\vec{\theta}$. 
Note that the computation of $\bra{b}\ket{\psi}$ is straightforward if $A$ is expressed as a linear combination of unitaries; see Sec.~\ref{sec:LCU}.
Another option that leads to the same result is~\cite{bravo:2019variational} 
\begin{align}
    \mathcal{L}_{\rm L}=1-\bra{\psi}BPB^\dagger\ket{\psi}~.
    \label{def:loss-VQLS}
\end{align}
Here, $\ket{b}=B\ket{0}$, and $P=\frac{1}{2}+\frac{1}{2n}\sum_{j=1}^nZ_j$ where $Z_j$ is the Pauli-$Z$ operator acting on the $j$-th qubit. 
These loss functions satisfy $\mathcal{L}_{\rm L}\le \mathcal{L}_{\rm G}\le n\mathcal{L}_{\rm L}$~\cite{bravo:2019variational}, and hence the minimization of $\mathcal{L}_{\rm L}$ is equivalent to that of $\mathcal{L}_{\rm G}$.

Compared to QPF-HHL, the variational formulation will usually require less circuit depth and will be simpler to implement, especially in the near-term hardware. Note that variational quantum algorithms are in the hybrid quantum-classical form: the loss function is evaluated using the quantum computer and quantum measurements, while the optimization for finding the optimized $\vec{\theta}$ is using classical algorithms. But the drawback is that usually those algorithms are only heuristic and cannot have rigorous guarantees, although, for some specific variational quantum ans\"{a}tze, it is shown that VQLS has comparable exponential speedup in the matrix size \cite{cerezo2021variational}.  

Specifically, for the power system setup, we could replace the HHL-assisted quantum power-flow algorithm with the variational version:
\begin{center}
\begin{algorithmic}
\State Algorithm: \textbf{QPF-VQLS}$(\mathbf{F}, \mathbf{U}_0, K_{\text{max}},\epsilon_{\text{inverse}}, , \epsilon_0)$
\\
\State Given $\mathbf{U}_0$ \Comment{Provide an initial guess}
\State Given $K_{\text{max}}$ \Comment{Define the maximum iteration number}
\State Given $k=0$ \Comment{Initialize the iteration index}
\State Given $\epsilon_0 >0$ \Comment{Accuracy tolerance}
\While{$k\le K_{\text{max}}$ or $|\mathbf{F}_k| \ge \epsilon_0$}
\State $\mathbf{F}_k\gets \mathbf{F}(\mathbf{U}_k)$ \Comment{Compute the mismatch}
\State $\mathbf{J}_k \gets \mathbf{J}(\mathbf{U}_k)= \frac{\partial \mathbf{F}(\mathbf{U})}{\partial \mathbf{U}} |_{\mathbf{U}=\mathbf{U}_k}$ \Comment{Update the Jacobian matrix}
\State Upload the matrix $\mathbf{J}_k$, and the state $\mathbf{F}_k$ to the quantum computer using Quantum Random Access Memory (QRAM).
\State $\ket{d\mathbf{U}_k} \gets -\ket{\mathbf{J}_k^{-1} \mathbf{F}_k}$ \Comment{Compute the increment of $\mathbf{U}$ using the VQLS algorithm \textbf{VQLS}$(\mathbf{J}_k,\mathbf{F}_k,\epsilon_{\text{inverse}})$ and download $\ket{d\mathbf{U}_k} =$\textbf{VQLS}$(\mathbf{J}_k,\mathbf{F}_k,\epsilon_{\text{inverse}})$ using the classical shadow.}
\State Determine $d\mathbf{U}_k$ from $\ket{d\mathbf{U}_k}$
\State $\mathbf{U}_{k+1} \gets \mathbf{U}_k  + d\mathbf{U}_k$ \Comment{Compute new $\mathbf{U}$}
\State $k \gets k+1$ \Comment{Update the index $k$}
\EndWhile
\State Output $\mathbf{U}_k$
\end{algorithmic}
\end{center}

\subsection{Variational Quantum Power Flow (VQPF)}\label{sec:VQPF}
Moreover, instead of considering using VQLS as the intermediate step, we could directly implement the full variational formulation of the power-flow problem. Note that the equation $F(\mathbf{U})=0$ could be written as
\begin{align}
\mathbf{U}^{\rm T}{{O}_{a}}\mathbf{U} ={{f}_{a}}~,~~~a=1,2,\ldots ,2{{N}_{\text{bus}}}~,
\end{align}
and hence
\begin{align}
\left\langle\mathbf{U}\right|{{O}_{a}}\left|\mathbf{U} \right\rangle =c^{-1}{{f}_{a}}~,~~~a=1,2,\ldots ,2{{N}_{\text{bus}}}~,
\end{align}
where $c=\mathbf{U}^{\rm T}\mathbf{U}$ and $O_a$s are sparse Hermitian matrices and $f_a$s are sparse vectors.\footnote{As used widely for the original HHL framework, we could transform non-Hermitian matrices $A$s to Hermitian matrices by adding one extra qubit. Namely, one could introduce
\begin{align}
\left( \begin{matrix}
   0 & A  \\
   {{A}^{\dagger }} & 0  \\
\end{matrix} \right)~.
\end{align}}
Because $c$ is an unknown number, we need to eliminate them from the equations. One way is simply to take a ratio and define the loss function as
\begin{align}
\mathcal{L}(\vec{\theta} )=\frac{1}{2}\sum\limits_{a=2}^{2N_\text{bus}}\left(\frac{\left\langle 0 \right|U^\dagger(\vec{\theta})O_a U(\vec{\theta})\left| 0 \right\rangle}{\left\langle 0 \right|U^\dagger(\vec{\theta})O_1 U(\vec{\theta})\left| 0 \right\rangle} -\frac{f_a}{f_1}\right)^2~.
\end{align}
Then, we could directly perform classical optimization algorithms to find the solution of $\vec{\theta}$ by using hybrid quantum-classical setups. The algorithm is formalized as the following, based on the gradient descent algorithm with the learning rate $\eta$,
\begin{align}
{{\theta }_{\ell }}(k+1)-{{\theta }_{\ell }}(k)=-\eta \left.\frac{\partial\mathcal{L}}{\partial\theta_\ell} \right|_{\vec{\theta} =\vec{\theta} (k)}~.
\end{align}
\begin{center}
\begin{algorithmic}
\State Algorithm: \textbf{VQPF}$(O_a, f_a, \vec{\theta}(0), K_{\text{max}}, , \epsilon_0)$
\\
\State Given $\vec{\theta}(0)$ \Comment{Provide an initial guess}
\State Given $K_{\text{max}}$ \Comment{Define the maximum iteration number}
\State Given $k=0$ \Comment{Initialize the iteration index}
\State Given $\epsilon_0 >0$ \Comment{Accuracy tolerance}
\While{$k\le K_{\text{max}}$ or $| \mathcal{L}| \ge \epsilon_0$}
\State Compute
$\left.\frac{\partial \mathcal{L}}{\partial \theta_\ell }\right|_{\vec{\theta} =\vec{\theta} (k)}$ \Comment{Compute the loss function gradient}
\State ${{\theta }_{\ell }}(k+1)-{{\theta }_{\ell }}(k)=-\eta \left.\frac{\partial \mathcal{L}}{\partial \theta_\ell }\right|_{\vec{\theta} =\vec{\theta} (k)}$. \Comment{Compute the update rule}
\EndWhile
\State Output $\vec{\theta}$ and download $\ket{\mathbf{U}}=U(\vec{\theta})\ket{0}$.
\end{algorithmic}
\end{center}

Using Variational Quantum Power Flow (VQPF), we do not need to recursively solve the equation using the Newtonian algorithm. Since the matrices and vectors are sparse, it is friendly for quantum devices to upload and download the results, thus friendly for near-term quantum devices because of their variational nature. However, we lose the connection between existing classical algorithms. Moreover, as far as we know, there is no theoretical argument to guarantee the computational advantage for such a quadratic optimization problem using variational algorithms. We leave further theoretical and numerical studies about VQPF in the future.    

Unlike QPF-HHL and QPF-VQLS, we download only the final result $\ket{\mathbf{U}}$. The cost to download is at most proportional to $N_{\rm bus}$, which is negligible compared to the cost of the classical algorithm. 

\subsection{Obtaining $\vec{x}$ from $\ket{x}$}\label{sec:normalization}
Finally, we briefly comment on normalization. In \textbf{QPF-HLL} and \textbf{QPF-VQLS}, we determine $\ket{x}=\frac{\sum_i x_i\ket{i}}{\sqrt{\sum_i |x_i|^2}}$, where  $\vec{x}=A^{-1}\vec{b}$. By downloading the  state, we obtain $\tilde{\vec{x}}\equiv\frac{\vec{x}}{|\vec{x}|}$, which is different from $\vec{x}$ by a normalization factor $|\vec{x}|=|A^{-1}\vec{b}|$. Because $\vec{b}$ is known,  we can determine the normalization factor by comparing $\vec{b}$ and $A\tilde{\vec{x}}=\frac{\vec{b}}{|\vec{x}|}$. This is a straightforward task on a classical device because $A$ is sparse and, furthermore, we do not have to compute all elements of $A\tilde{\vec{x}}$. If the solution is exact, we only need one nonzero element, and hence, the computational cost does not scale with the size of the matrices.

In \textbf{VQPF}, the normalization factor $c$ can be determined by comparing $\left\langle\mathbf{U}\right|{{O}_{a}}\left|\mathbf{U} \right\rangle$, which can be calculated on a quantum device, with $f_a$. 
\section{Requirements for quantum random access memory}\label{qram_require}
Quantum Random Access Memory (QRAM) plays an important role in the algorithms presented in this paper. In this section, we relate the size of the problem to the QRAM hyperparameters. 

For instance, in our algorithm \textbf{QPF-HHL}, $\mathbf{F}_k$ is uploaded as a quantum state $\ket{\mathbf{F}_k}$. 
The data-lookup oracle \eqref{data-lookup-oracle} is used. It requires $\mathcal{O}(\log N_{\rm bus})$ switches with $\mathcal{O}(N_{\rm bus})$ qubits. Moreover, as a part of the HHL algorithm, Quantum Phase Estimation (QPE) is used. QPE uses the Hamiltonian time evolution regarding $\mathbf{J}_k$ as Hamiltonian, and QRAM enables us to use the efficient implementation of the Hamiltonian time evolution. More specifically, two kinds of oracles of the following forms are constructed by using QRAM:
\begin{align}
    &
    O_{\mathbf{J}_k}^{\rm (DL)}\ket{p,q}\ket{0}=\ket{p,q}\ket{\mathbf{J}_{k,pq}}\nonumber\\
    &
    O_f\ket{p}\ket{q}=\ket{p}\ket{f(p,q)}. 
\end{align}
Here $O_{\mathbf{J}_k}^{\rm (DL)}$ is a data-lookup oracle that returns the $(p,q)$-element of matrix $\mathbf{J}_k$ for any $p,q=1,2,\cdots,2N_{\rm bus}$. 
Te second oracle $O_f$ specifies the location of the nonzero entries, i.e., $f(p,q)$ is the $q$-th non-zero entry in the $p$-th row. These oracles require $\mathcal{O}(\log (sN_{\rm bus}))$ switches with $\mathcal{O}(sN_{\rm bus})$ qubits, where $s$ is the sparsity (number of nonzero entries at each row). 

It is challenging to make physical realizations of QRAM. Here, we make a theoretical estimate of the required QRAM resources for the quantum power-flow problem. 

\begin{figure}[ht]
\centering
\includegraphics[width=1.00\textwidth]{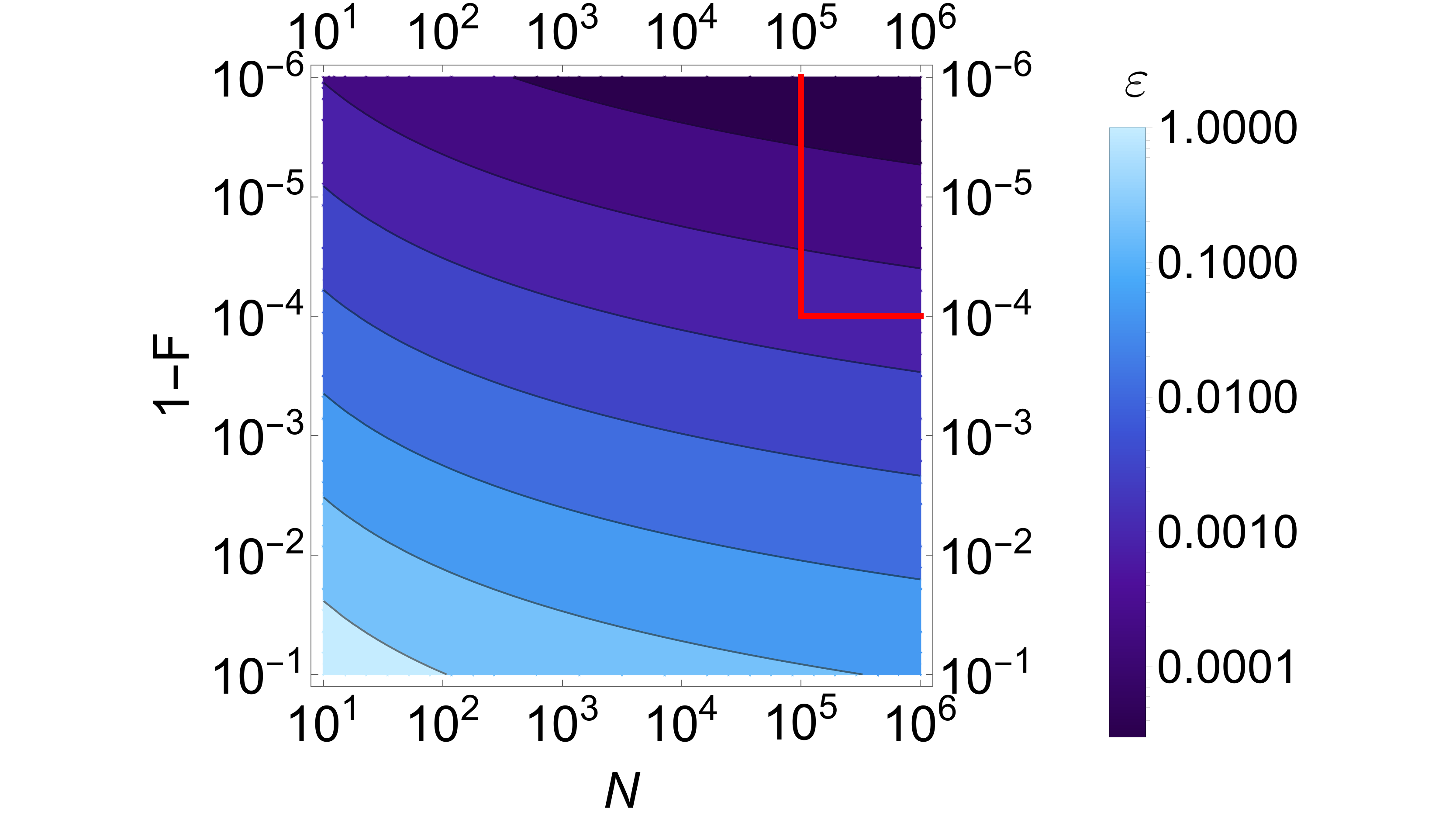}
\caption{Bounds on the error rate $\varepsilon$ from quantum power-flow problem. Here we compute the required error rate from the precision of the problem $1-F$ and the size of the data $N$ assuming the time efficiency $T = \log N$. The region inside the red box is the state-of-the-art requirement for modern power flows in the smart grid.
}
\label{fig:benchmark1}
\end{figure}

From \cite{hann2021resilience}, it is proven that for the bucket-brigade QRAM architectures~\cite{giovannetti2008quantum}, one of the primary methods for realizing QRAM unitaries, the infidelity $1-F$ will scale the size of the data $N$ as
\begin{align}
    1-F \sim \frac{1}{4} \varepsilon T \log N \sim \frac{1}{4} \varepsilon \log^2 N ~.
\end{align}
Here, $T=\mathcal{O} (\log N)$ is the time required to perform a query, and $\varepsilon$ is the probability of error per time step. Consider that in the modern power system problem, we require computations for the size of the matrix $N \sim 2N_{\rm{bus}}\sim 10^5$, and the precision $1-F \sim 10^{-4}$. In Figure \ref{fig:benchmark1}, we estimate the constraints from the modern power system problem towards the error rate $\varepsilon$ on the hardware. 

\begin{figure}[ht]
\centering
\includegraphics[width=1.00\textwidth]{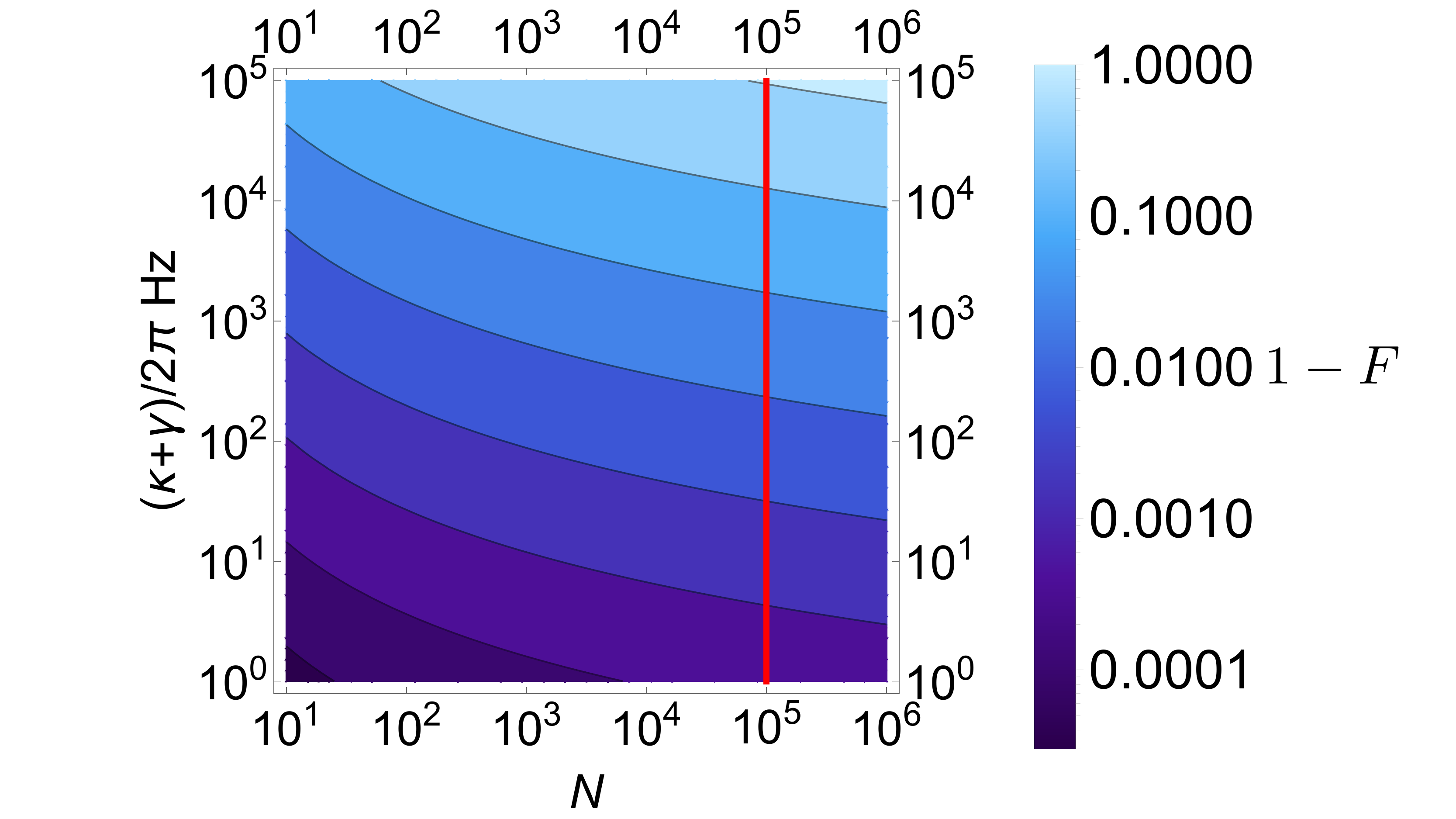}
\caption{Bounds on the precision $1-F$ of the quantum power-flow problem from physical decoherence rate $\kappa+\gamma$. Here we assume $g_d=1 \text{ kHz} \times 2\pi$, $\nu = 10 \text{ MHz} \times 2\pi$, and $c_d =4.5$ which is the average of the CZ and SWAP gates inside the QRAM circuit as an estimate. The red line emphasizes the typical matrix scales used in the modern power-flow problem in the smart grid study.}
\label{fig:benchmark2}
\end{figure}

One could estimate the error rate $\varepsilon$ based on specific hardware realizations of QRAM. For instance, \cite{PhysRevLett.123.250501} describes a specific hardware realization of QRAM based on the hybrid quantum acoustic systems, which is shown to be hardware-efficient and noise-resilient \cite{hann2021resilience}. The design of QRAM circuits is established with the help of the circuit quantum acoustodynamics. The transmon qubit is coupled to several acoustic modes piezoelectrically, supported by bulk acoustic wave resonators, surface acoustic wave resonators, or the phononic crystal resonators arrays (see \cite{PhysRevLett.123.250501} in more details). The effective Hamiltonian is given by,
\begin{align}
H &=\omega_{q} q^{\dagger} q-\frac{\alpha}{2} q^{\dagger} q^{\dagger} q q +\sum_{k}\left(\omega_{k} m_{k}^{\dagger} m_{k}+g_{k} q^{\dagger} m_{k}+g_{k}^{*} q m_{k}^{\dagger}\right)+H_{d}~,
\end{align}
with the transmon mode $q$ with energy $\omega_q$, phonon mode $m_k$ with energy $\omega_k$, the Kerr non-linearity $\alpha$, the coupling strength $g_k$, and the external drives of the transmon $H_d$. One could estimate $\varepsilon$ by the infidelity of the direct gate per query, 
\begin{align}
\varepsilon \approx (\kappa +\gamma )\frac{{{c}_{d}}\pi }{2{{g}_{d}}}+{{\left( \frac{{{g}_{d}}}{\nu } \right)}^{2}}~.
\end{align}
Here, $\kappa$ and $\gamma$ are bare phonon- and transmon-decoherence rates, $g_d$ is the direct coupling, $\nu$ is the free spectral range, and $c_d = \mathcal{O} (1)$ is the constant when computing the duration of each gate. We estimate the constraints for the required decoherence rate $\kappa +\gamma$ from given quantum power-flow problems in Figure \ref{fig:benchmark2}.

Note that in the long term, fault-tolerant quantum computing is believed to be required when performing the HHL algorithm. Thus, the QRAM architecture is required to be fault-tolerant as well. There exists an estimate about realizing QRAM fault-tolerance on a small scale \cite{arunachalam2015robustness}. However, it is challenging to realize quantum error correction in the QRAM circuits for large-scale databases \cite{liu2022quantum}.

\section{Toward implementation on quantum devices}
In this section, we discuss several technical issues necessary for the implementation of the algorithms proposed in this paper on quantum devices. 
In Sec.~\ref{sec:LCU}, we introduce the LCU decomposition.  
In Sec.~\ref{sec:QPF-HHL-resource-counting}, we estimate the resource needed for QPF-HHL. Not surprisingly, a large amount of resources is needed and it is unlikely that QPF-HHL can be used in the near-term devices. 
In Sec.~\ref{sec:experiments}, we run the VQLS algorithm which is a subroutine of QPF-VQLS.

\subsection{LCU decomposition}\label{sec:LCU}
It is crucial to inform the quantum computer what the matrix elements of $\mathbf{J}_k$ are before we run the quantum algorithm. An efficient way of uploading matrix elements is the so-called Linear Combinations of Unitaries
(LCU), which could be implemented using QRAMs.

\begin{figure}[ht]
\centering
\includegraphics[width=1.00\textwidth]{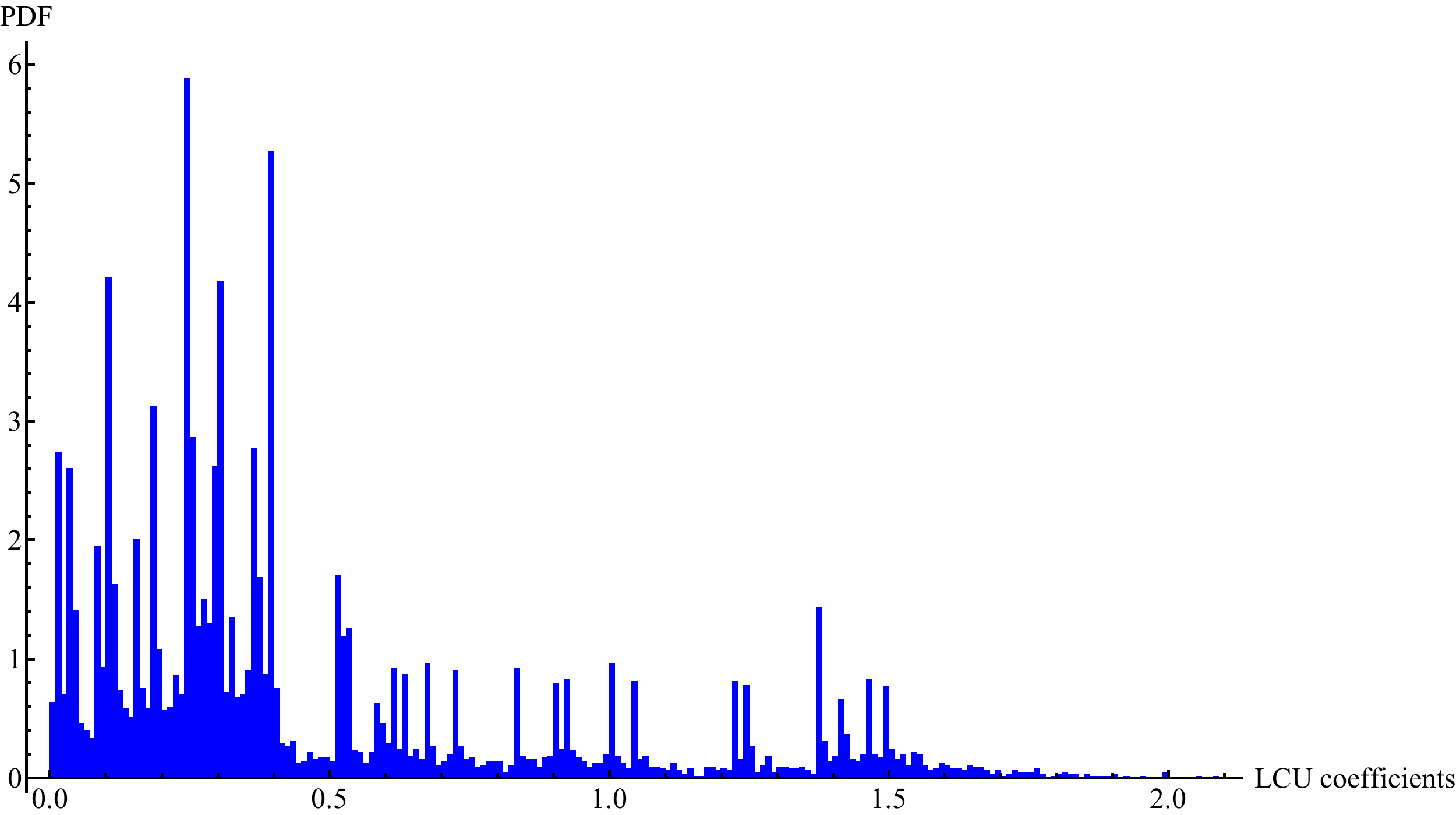}
\caption{An example of the statistics of the LCU coefficients in the power-flow problem. Here we have 102 different $64 \times 64$ Hermitian matrices from the real power-flow problem. In this case, we have $835\pm 8$ non-zero coefficients in the LCU decomposition. 
The Probability Distribution Function (PDF) of the nonzero coefficients is shown. 
}
\label{fig:lcu}
\end{figure}

For a given Hermitian matrix $A$, one can decompose it by
\begin{align}
A = \sum_i a_i P_i~, 
\qquad a_i\in\mathbb{R}~.
\end{align}
Here, $P_i$s are Hermitian and \emph{unitary} at the same time. A practical choice of the set $P_i$ is from the Pauli group. The $n$-qubit Pauli group $\mathcal{P}$ is defined as 
\begin{align}
\mathcal{P}=\left\{ \otimes _{q=1}^{n}{{p}_{q}}; {{p}_{q}}= I,X,Y,\ {\rm or}\ Z \right\}~.
\end{align}
$\mathcal{P}$ is a group under the operation of the matrix product, modulo the phase in front of the definition of Pauli matrices. In total, $\mathcal{P}$ has $4^n$ elements. 
From here on, we use $P_i$ ($i=1,2,\cdots,4^n$) to denote the elements of $\mathcal{P}$. 
One can decompose an arbitrary matrix with the $2^n \times 2^n =4^n$ degrees of freedom into the $n$-qubit Pauli group basis. One could determine the coefficients $a_i$ by using 
\begin{align}
    {{a}_{i}}=\frac{1}{{{2}^{n}}}\operatorname{Tr}\left( {{P}_{i}}A \right)~.
\end{align}
One could determine the coefficients $a_i$ and understand the complexity of the LCU oracle in our practical power-flow setup. In Figure \ref{fig:lcu}, we provide the probability distribution of the nonzero coefficients in an example of $64\times 64$ matrices $\mathbf{J}_k$ ($k=1,2,\cdots,102$) obtained from the power-flow problem. (This example is studied in Sec.~\ref{sec:experiments}.)
There are $64\times 64=4096$ terms in the LCU decomposition. Among them $835\pm 8$ coefficients are nonzero. 

There are alternative choices of the basis instead of Pauli group elements. In \cite{lapworth2022hybrid}, alternative schemes using Hadamard products have been proposed. We leave further analysis about the optimal basis for power-flow problems in future research.

\subsection{Resource counting for QPF-HHL}\label{sec:QPF-HHL-resource-counting}
In our setup, the HHL algorithm consists of the following subroutines (see also Section \ref{sec:QPF-HHL}): 
\begin{itemize}
\item Quantum Phase Estimation (QPE). 
\item Eigenvalue inversions. 
\item Inverse Quantum Phase Estimation (IQPE).  
\item Hamiltonian simulation via LCU decomposition. This subroutine is used in QPE and IQPE. 
\end{itemize}
Note that in the eigenvalue inversion steps, e.g. in Equation \eqref{eq:hhl-inversion}, it is customary to perform another subroutine: oblivious amplitude amplification, to boost the probability of measuring $\ket{1}$ close to $1$ in $O(\frac{\lambda_i }{C})$ rounds so that it would effectively reduce the number of queries needed to achieving correct ancilla measurement. In our numerical simulation, however, we discard counting any query complexity but rather the circuit depth in assembling one instance of the QPF-HHL circuit; and as a result, we skip the implementation of the oblivious amplitude amplification.  

\begin{figure}[ht]
\centering
\includegraphics[width=0.8\textwidth]{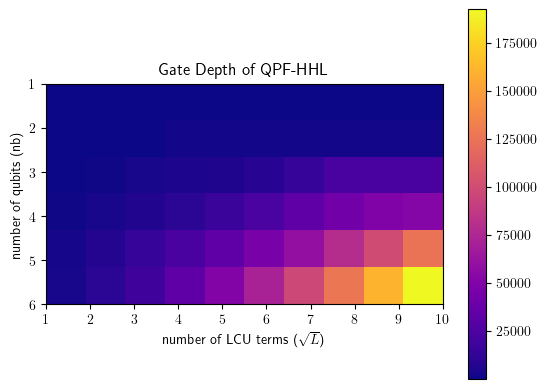}
\caption{An example of the circuit depth running the QPF-HHL algorithms. The depth of circuits scale with the increasing number of qubits and number of LCU terms in square root. Since every Pauli basis is 1-sparse, the square of the row also gives the sparsity reading of the Jacobian matrix. $M$ is taken to be 10. }
\label{fig:opf-hhl}
\end{figure}
The Hamiltonian simulation utilizes standard Trotterization. Given a Hamiltonian written in Pauli basis $\{P_i\}$ as
\begin{align}
    H = \sum^L_{i=1} a_i P_i, 
\end{align}
where $L$ is the number of nonzero coefficients. In the numerical simulation, the Hamiltonian is the Jacobian matrices ranging from dimension $2 \times 2 $ to $64 \times 64$ from real power-flow data. Instead of $U(t) \equiv \exp(itH)$ we simulate 
\begin{align}
    \widetilde{U}(t) = \left(\Pi^L_{i=1} \exp(it/M a_iP_i)\right)^M
\end{align}
Where $M = O(t^2/\epsilon)$ so that $\| U(t) - \widetilde{U}(t) \| \leq \epsilon $. 
\begin{tiny}
\begin{table}[h!]
\centering
\begin{tabular}{|c c c c c c c c c c c c|} 
\hline
& $2^n|L$ & 1&  4   & 9    & 16   & 25     & 36 & 49 & 64 & 81 & 100  \\
\hline
&  $2$ & 75 &  ... &  ... &  ... &  ... &  ... & ... & ... & ...& ...\\
\hline
 &$4$ & 273  & 274 &  575 &  2215 &  ... &  ... &  ... &...&...&... \\
 \hline
 & $8$ & 613  & 1500 &  3110 &  5086 &  5970 &  8853 &  15048 & 22922 &...& ... \\
\hline
&  $16$ & 1161 &  3090 &  6466 &  10071 &  16165 &  23831 &33136&42836&49425&52547\\
  \hline
  & $32$ & 2663 &  7309 &  14553 &  23229 &  33493 &  45654 &58951 & 79403 &99607 &124791\\
\hline
&  $64$ & 3542 &  9845 &  18376 &  33799 &  50641 &  719064 &97064 &126624 &160138 &192351\\
\hline
\end{tabular}
\caption{The gate depth of \textbf{QPF-HHL} circuit for Jacobian matrix with different dimensions $2^n$ and sparsity pattern as counted by the number of LCU terms $L$. Entry with $\dots$ implies that the given Jacobian has LCU terms less than the number indicated by the column. }
\label{table:1}
\end{table}
\end{tiny}

\begin{figure}[ht]
\centering
\includegraphics[width=0.7\textwidth]{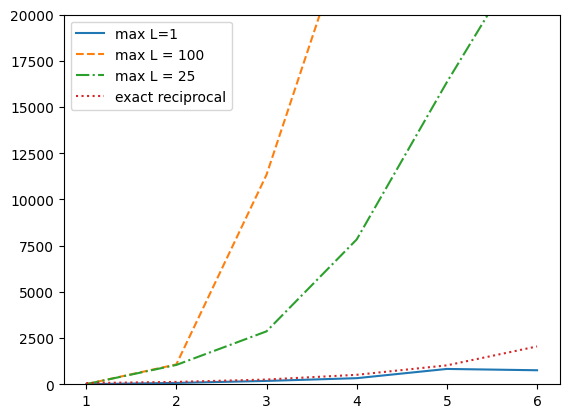}
\caption{Schematic plot showing the resource taken by QPE with different upper bounds on $L$ compared with that by exact eigenvalue inversion Equation \eqref{eq: reciporcal} with $p(x_i) = \arcsin{\frac{C}{x_i}}$. The vertical and horizontal axes are the gate depth and the number of qubits $n$, respectively. When $4^n<L$, the upper bound is set to $4^n$ instead of $L$.
The gate depth is close to half of that of the QPF-HHL circuit, i.e., QPE and IQPE give the dominant contribution to the gate countings in QPF-HHL.
Due to a large overall coefficient, more gates are required than a naive implementation of the eigenvalue inversion whose gate counting scales exponentially. 
}
\label{fig: qpe-depth}
\end{figure}

In Table~\ref{table:1}, Figure~\ref{fig:opf-hhl}, and Figure~\ref{fig: qpe-depth}, it is shown how the gate depth of the QPF-HHL circuit depends on $n$ and $L$. In the circuit simulation using IBM qiskit \cite{qiskit}, we set $M=10$. 
By definition, $L\le 4^n$. In Figure~\ref{fig:opf-hhl} and Table~\ref{table:1}, the upper bound is set to $4^n$ instead of $L$ when $4^n<L$. 
The quick growth of gate counting is primarily due to the large overhead performing the QPE and IQPE. The QPE and IQPE, which are essentially the same thing, are the key subroutines to our implementation of the HHL algorithm.
The gate counting for the HHL in Table~\ref{table:1} and Figure~\ref{fig:opf-hhl} are approximately twice as large as the gate counting for QPE shown in Figure~\ref{fig: qpe-depth}. 
This means that QPE and IQPE practically dominate the cost. 
Furthermore, because of a large overall coefficient, they require more cost than a naive implementation of the eigenvalue inversion, whose gate counting scales exponentially, when $n$ is small; see Figure~\ref{fig: qpe-depth} and the next paragraph.
Therefore, to enjoy the benefit from the exponential speedup, we should improve or replace the QPE subroutine in the HHL algorithm.  
A potential option is to avoid the use of QPE following Ref~\cite{childs2017quantum} and improve HHL circuit complexity from $\operatorname{poly}(n, 1/\epsilon)$ to $\operatorname{poly}(n, \log (1/\epsilon))$.

The eigenvalue inversion \eqref{eigenvalue-inversion} implements the following controlled rotation: 
\begin{align}\label{eq: reciporcal}
   P_p: \ket{\lambda}\ket{0} \mapsto \ket{\lambda}\left( \cos (p(\lambda)) \ket{0} + \sin (p(\lambda))\ket{1}\right)~, 
\end{align}
where $p(\lambda) = \arcsin(\frac{C}{\lambda})$.
A naive, exact implementation of this requires $2^n -1$ conditional rotation gates. 
More generally, $p(\lambda)$ could be approximated by polynomials of degree $d$ so that $P_p$ requires $O(n^d)$-number of CNOT and single-qubit gates~\cite{woerner2019quantum}. This polynomial scaling is favorable in computing large matrices.
However, in case that matrix size and the number of qubits are at most $64 \times 64$ and $n=7$ (since the matrix eigenvalue could be negative so additional qubit is needed to store the sign), the exponential scaling of a naive implementation is buried into the large amount of resources required for performing QPE. We need to worry about the exponential scaling in practice when the Jacobian matrix $F$ is local and sparse. 

In future works toward large-scale implementation, we might consider utilizing polynomial state preparation (PSP) \cite{woerner2019quantum} using a truncated Taylor series polynomial or the piecewise Chebyshev polynomial proposed in \cite{vazquez2022enhancing} to ensure an efficient implementation of the eigenvalue inversion.

\subsection{VQLS implementation: simulation and experiments with 6 qubits}\label{sec:experiments}
\begin{figure}[hbt!]
\centering
\includegraphics[width=0.8\textwidth]{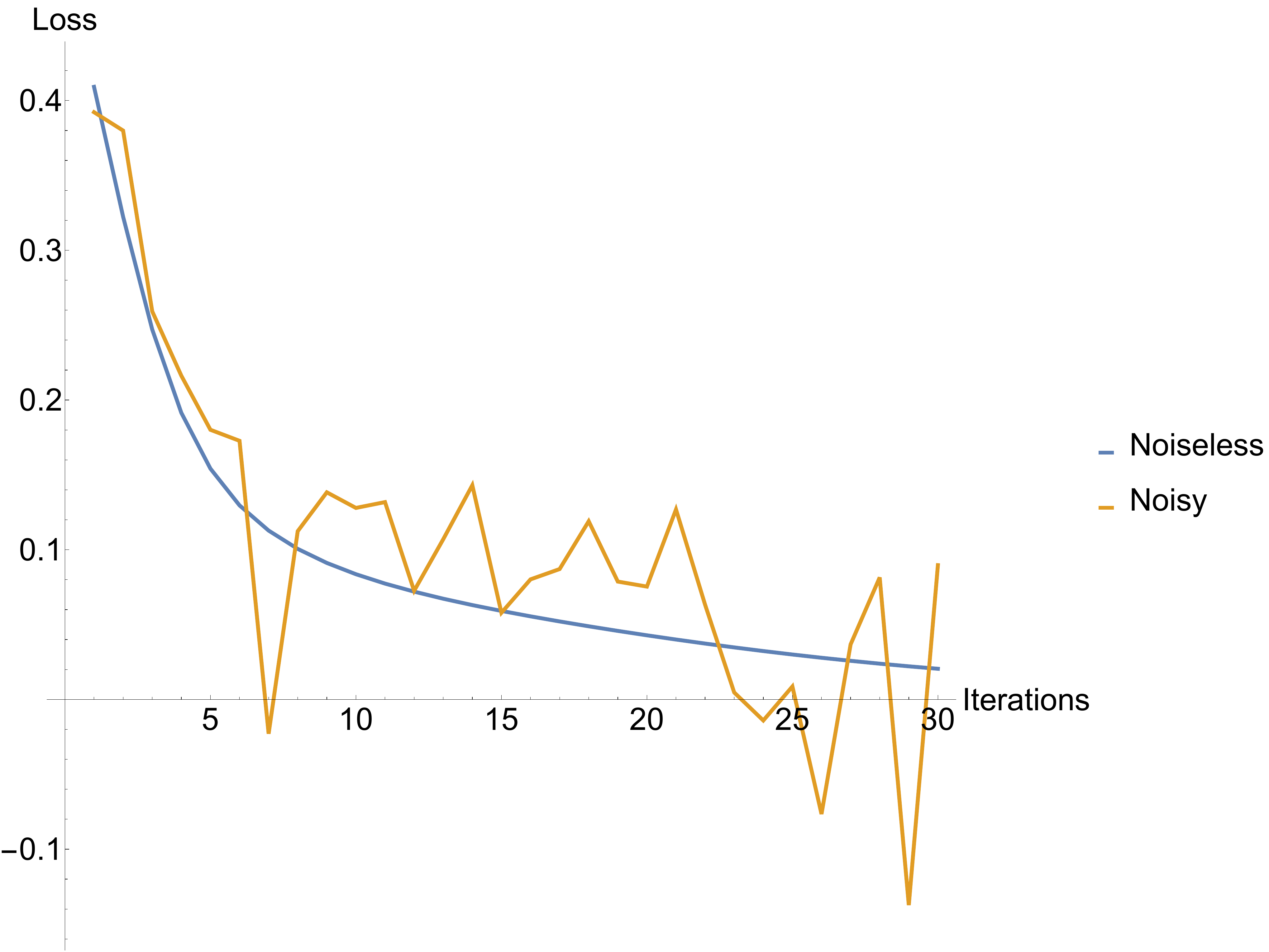}
\includegraphics[width=0.7\textwidth]{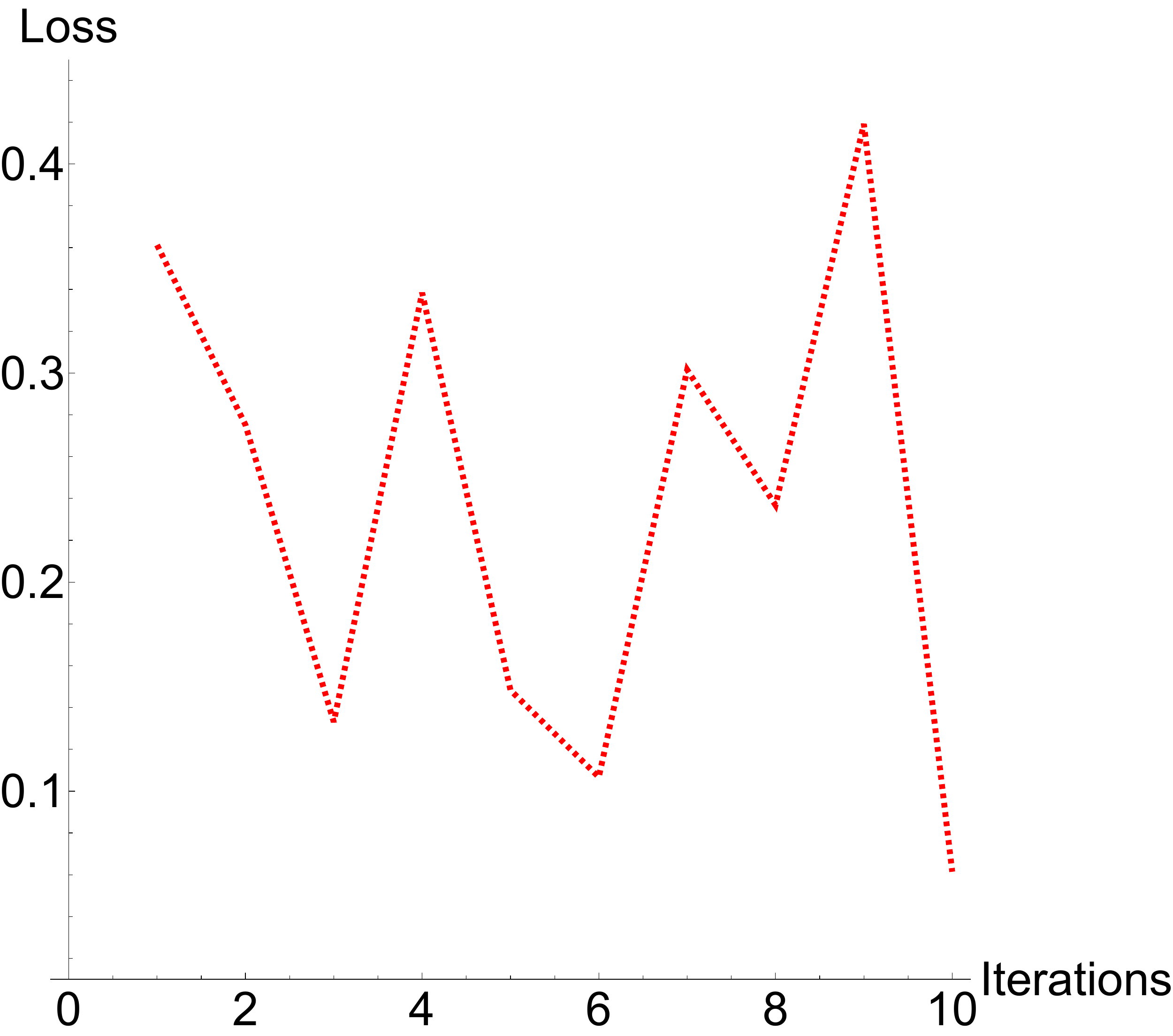}
\caption{ VQLS simulation of real power-flow problems in a single instance. 
[Top] Simulations on a classical device with and without noise. 
The noisy simulation is from the noise model of \texttt{ibm\_nairobi}.
[Bottom] VQLS real experiment of real power-flow problems in a single instance. Here the experiment is from \texttt{ibmq\_jakarta}.
The loss function is $\mathcal{L}_{\rm L}$ defined by \eqref{def:loss-VQLS}. 
Although the loss function is non-negative by definition, negative values may appear due to the feature of the noise model. 
}
\label{fig:variational}
\end{figure}

We run the VQLS algorithm by considering the variational ansatz used in \cite{bravo:2019variational}. 
We consider $N_{\rm bus}=14$, to which the $2N_{\rm bus}=28 \approx 2^5$, and extend it to a Hermitian matrix by doubling the size to $2^6$ -component vector $\mathbf{U}$ can be realized by using 6 qubits. 
We choose one typical matrix from the 102 realizations of the $64\times 64$ Hermitian matrices from the 14-bus standard power-flow problem (in Figure \ref{fig:lcu}) and include the first few non-trivial LCU coefficients. (On noisy devices, we cannot include many LCU coefficients.) 
The first panel in Figure~\ref{fig:variational} shows the result of our dynamical optimization process in the noiseless and noisy cases for our VQLS in a typical step, where the noisy simulation is performed using the noise model of \texttt{ibm\_nairobi} from the real data with 5 truncated LCU coefficients and 2 layers in the variational circuits. 
The second panel in Figure \ref{fig:variational} shows the real quantum experiment with \texttt{ibmq\_jakarta} with 3 truncated LCU coefficients and 1 layer in the variational circuits. We notice that in our 6-qubit case, the variational simulation could still reasonably converge in the noiseless setting, but strongly fluctuate in the noisy setting and the real device cases. However, we get final infidelity around $\mathcal{O}(5\%)\sim \mathcal{O}(10\%)$ in those cases, even in the real machine with 10 iterations.


The results of noisy simulations show the potential limitations on the scale of VQLS simulation because of quantum noises. On the other hand, more careful implementations and error mitigation might still be possible to reduce part of the noise, and we will leave more detailed investigations in the future. The simulation is powered using qBraid system~\cite{qbraid} and IBM Qiskit~\cite{qiskit}. 

\begin{figure}[hbt!]
\centering
\includegraphics[width=0.9\textwidth]{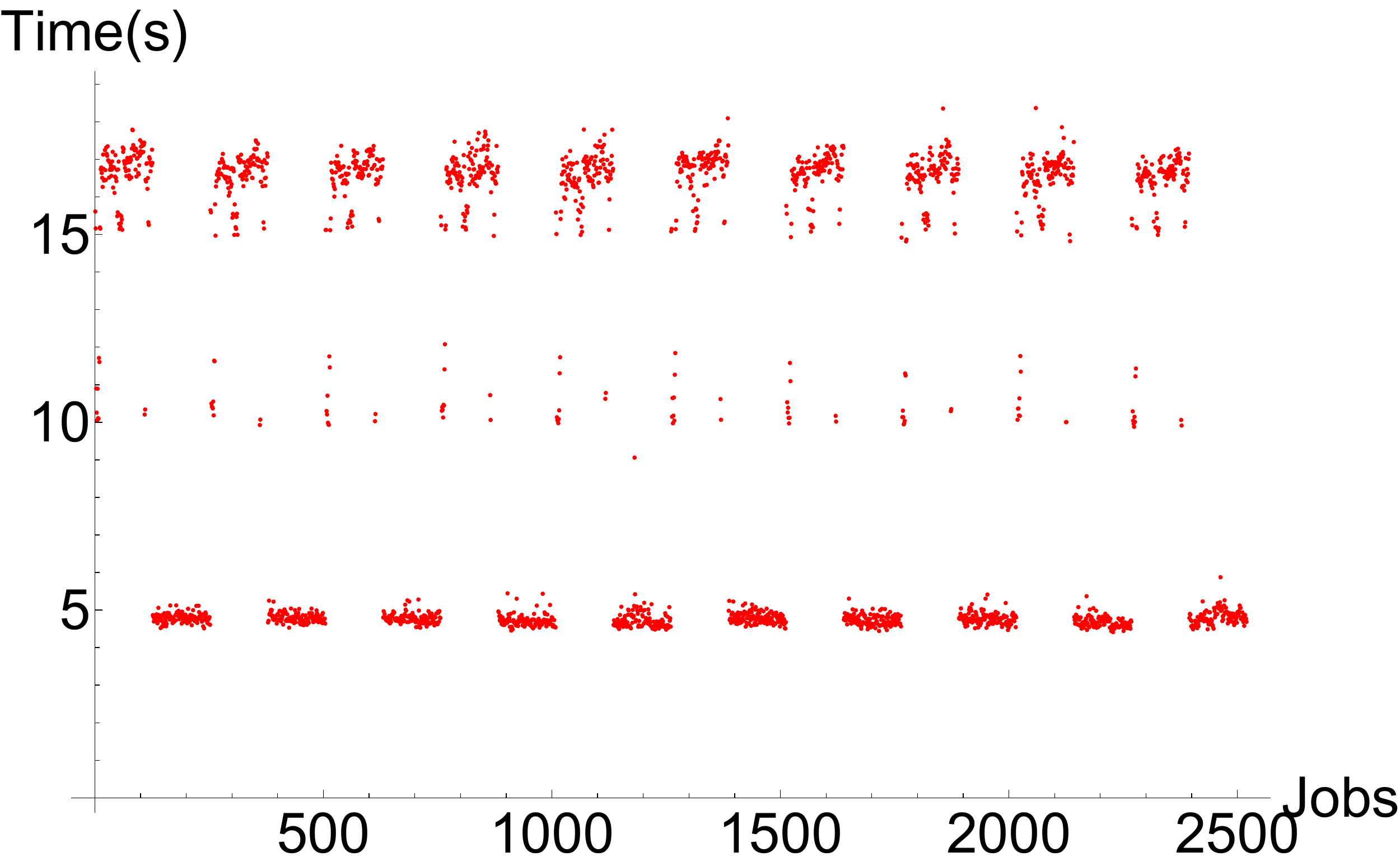}
\caption{The distribution of the running time in the real quantum experiment. The horizontal axis is the label of jobs (with 2520 in total), while the vertical axis is the time one spends on each job (measured in seconds). Each job runs 10.4 seconds on average, and in total, 7.3 hours.
The same pattern repeats 10 times because we had 10 steps for the VQSL. 
}
\label{fig:time}
\end{figure}

For real simulations, the duration of each job on a quantum device is shown in Figure \ref{fig:time}. We use 10 iterations in total with 2520 jobs in \texttt{ibmq\_jakarta}, with 26297 seconds (7.3 hours) in total.
\section*{Acknoowledgement}
We thank Liang Jiang for the discussions. JL is supported in part by International Business Machines (IBM) Quantum through the
Chicago Quantum Exchange, and the Pritzker School
of Molecular Engineering at the University of Chicago
through AFOSR MURI (FA9550-21-1-0209). MH thanks the STFC Ernest Rutherford
Grant ST/R003599/1 and the Royal Society International Exchanges award IEC/R3/213026. DW, supported in part by DOE SETO DE-EE0009031 and NSF CNS-1735463, thanks Dr. Marija Ilic at LIDS, MIT for the helpful discussions and support.

\bibliographystyle{utphys}
\bibliography{References.bib}

\providecommand{\href}[2]{#2}\begingroup\raggedright\begin{thebibliography}{10}

\bibitem{manabe:1975effects}
S.~Manabe and R.~T. Wetherald, ``The effects of doubling the co 2 concentration
  on the climate of a general circulation model,'' {\em Journal of Atmospheric
  Sciences} {\bfseries 32} no.~1, (1975) 3--15.

\bibitem{manabe:1991transient}
S.~Manabe, R.~J. Stouffer, M.~J. Spelman, and K.~Bryan, ``Transient responses
  of a coupled ocean--atmosphere model to gradual changes of atmospheric co2.
  part i. annual mean response,'' {\em Journal of Climate} {\bfseries 4} no.~8,
  (1991) 785--818.

\bibitem{manabe:1992transient}
S.~Manabe, M.~Spelman, and R.~Stouffer, ``Transient responses of a coupled
  ocean-atmosphere model to gradual changes of atmospheric co 2. part ii:
  Seasonal response,'' {\em Journal of Climate} {\bfseries 5} no.~2, (1992)
  105--126.

\bibitem{masson:2018global}
V.~Masson-Delmotte, P.~Zhai, H.-O. P{\"o}rtner, D.~Roberts, J.~Skea, P.~R.
  Shukla, A.~Pirani, W.~Moufouma-Okia, C.~P{\'e}an, R.~Pidcock, {\em et~al.},
  ``Global warming of 1.5 c,'' {\em An IPCC Special Report on the impacts of
  global warming of} {\bfseries 1} no.~5, (2018) .

\bibitem{epa}
U.~S. E.~P. Agency, ``Sources of greenhouse gas emissions.'' [online].
  available, https://www.epa.gov/ghgemissions/sources-greenhouse-gas-emissions.

\bibitem{nerc:2017}
N.~A. E.~R. Corporation, ``1200 mw fault induced solar photovoltaic resource
  interruption disturbance report.'' [online]. available,
  https://www.nerc.com/pa/rrm/ea/pages/1200-mw-fault-induced-solar-photovoltaic-resource-interruption-disturbance-report.aspx.

\bibitem{eso:2020}
N.~G. ESO, ``Information about the 9 august power cut and the eso.'' [online].
  available,
  https://www.nationalgrideso.com/information-about-great-britains-energy-system-and-electricity-system-operator-eso.

\bibitem{ward:1956digital}
J.~Ward and H.~Hale, ``Digital computer solution of power-flow problems
  [includes discussion],'' {\em Transactions of the American Institute of
  Electrical Engineers. Part III: Power Apparatus and Systems} {\bfseries 75}
  no.~3, (1956) 398--404.

\bibitem{mcgillis:1957nodal}
D.~McGillis, ``Nodal iterative solution of power-flow problem using ibm 604
  digital computer,'' {\em Transactions of the American Institute of Electrical
  Engineers. Part III: Power Apparatus and Systems} {\bfseries 76} no.~3,
  (1957) 803--809.

\bibitem{jordan:1957rapidly}
R.~Jordan, ``Rapidly converging digital load flow,'' {\em Transactions of the
  American Institute of Electrical Engineers. Part III: Power Apparatus and
  Systems} {\bfseries 76} no.~3, (1957) 1433--1438.

\bibitem{james:1961}
J.~E. Van~Ness, ``Elimination methods for load-flow studies,'' {\em
  Transactions of the American Institute of Electrical Engineers. Part III:
  Power Apparatus and Systems} {\bfseries 80} no.~3, (1961) 299--302.

\bibitem{tinney:1967power}
W.~F. Tinney and C.~E. Hart, ``Power flow solution by newton's method,'' {\em
  IEEE Transactions on Power Apparatus and systems} no.~11, (1967) 1449--1460.

\bibitem{iwamoto:1978fast}
S.~Iwamoto and Y.~Tamura, ``A fast load flow method retaining nonlinearity,''
  {\em IEEE Transactions on Power Apparatus and Systems} no.~5, (1978)
  1586--1599.

\bibitem{trias:2012holomorphic}
A.~Trias, ``The holomorphic embedding load flow method,'' in {\em 2012 IEEE
  Power and Energy Society General Meeting}, pp.~1--8, IEEE.
\newblock 2012.

\bibitem{chiang:2017novel}
H.-D. Chiang, T.~Wang, and H.~Sheng, ``A novel fast and flexible holomorphic
  embedding power flow method,'' {\em IEEE Transactions on Power Systems}
  {\bfseries 33} no.~3, (2017) 2551--2562.

\bibitem{wu:2019holomorphic}
D.~Wu and B.~Wang, ``Holomorphic embedding based continuation method for
  identifying multiple power flow solutions,'' {\em IEEE access} {\bfseries 7}
  (2019) 86843--86853.

\bibitem{wang:2020theoretical}
T.~Wang and H.-D. Chiang, ``Theoretical study of non-iterative holomorphic
  embedding methods for solving nonlinear power flow equations: Algebraic
  property,'' {\em IEEE Transactions on Power Systems} {\bfseries 36} no.~4,
  (2020) 2934--2945.

\bibitem{eskandarpour:2020quantum}
R.~Eskandarpour, P.~Gokhale, A.~Khodaei, F.~T. Chong, A.~Passo, and
  S.~Bahramirad, ``Quantum computing for enhancing grid security,'' {\em IEEE
  Transactions on Power Systems} {\bfseries 35} no.~5, (2020) 4135--4137.

\bibitem{feng:2021quantum}
F.~Feng, Y.~Zhou, and P.~Zhang, ``Quantum computing for enhancing grid
  security,'' {\em IEEE Transactions on Power Systems} {\bfseries 36} no.~4,
  (2021) 3810--3812.

\bibitem{zhou:2021quantum}
Y.~Zhou, F.~Feng, and P.~Zhang, ``Quantum electromagnetic transients program,''
  {\em IEEE Transactions on Power Systems} {\bfseries 36} no.~4, (2021)
  3813--3816.

\bibitem{saevarsson:2022quantum}
B.~S{\ae}varsson, S.~Chatzivasileiadis, H.~J{\'o}hannsson, and
  J.~{\O}stergaard, ``Quantum computing for power flow algorithms: Testing on
  real quantum computers,'' {\em arXiv preprint arXiv:2204.14028} (2022) .

\bibitem{eskandarpour:2021experimental}
R.~Eskandarpour, K.~Ghosh, A.~Khodaei, and A.~Paaso, ``Experimental quantum
  computing to solve network dc power flow problem,'' {\em arXiv preprint
  arXiv:2106.12032} (2021) .

\bibitem{harrow:2009quantum}
A.~W. Harrow, A.~Hassidim, and S.~Lloyd, ``Quantum algorithm for linear systems
  of equations,'' {\em Physical Review Letters} {\bfseries 103} no.~15, (2009)
  150502.

\bibitem{giovannetti2008quantum}
V.~Giovannetti, S.~Lloyd, and L.~Maccone, ``Quantum random access memory,''
  {\em Physical review letters} {\bfseries 100} no.~16, (2008) 160501.

\bibitem{hann2021practicality}
C.~T. Hann, {\em Practicality of Quantum Random Access Memory}.
\newblock PhD thesis, Yale University, 2021.

\bibitem{hann2021resilience}
C.~T. Hann, G.~Lee, S.~Girvin, and L.~Jiang, ``Resilience of quantum random
  access memory to generic noise,'' {\em PRX Quantum} {\bfseries 2} no.~2,
  (2021) 020311.

\bibitem{huang2020predicting}
H.-Y. Huang, R.~Kueng, and J.~Preskill, ``Predicting many properties of a
  quantum system from very few measurements,'' {\em Nature Physics} {\bfseries
  16} no.~10, (2020) 1050--1057.

\bibitem{bravo:2019variational}
C.~Bravo-Prieto, R.~LaRose, M.~Cerezo, Y.~Subasi, L.~Cincio, and P.~J. Coles,
  ``Variational quantum linear solver,'' {\em arXiv preprint arXiv:1909.05820}
  (2019) .

\bibitem{cerezo2021variational}
M.~Cerezo, A.~Arrasmith, R.~Babbush, S.~C. Benjamin, S.~Endo, K.~Fujii, J.~R.
  McClean, K.~Mitarai, X.~Yuan, L.~Cincio, {\em et~al.}, ``Variational quantum
  algorithms,'' {\em Nature Reviews Physics} {\bfseries 3} no.~9, (2021)
  625--644.

\bibitem{farhi2018classification}
E.~Farhi and H.~Neven, ``Classification with quantum neural networks on near
  term processors,'' {\em arXiv preprint arXiv:1802.06002} (2018) .

\bibitem{PhysRevLett.123.250501}
C.~T. Hann, C.-L. Zou, Y.~Zhang, Y.~Chu, R.~J. Schoelkopf, S.~M. Girvin, and
  L.~Jiang, ``Hardware-efficient quantum random access memory with hybrid
  quantum acoustic systems,''
  \href{http://dx.doi.org/10.1103/PhysRevLett.123.250501}{{\em Phys. Rev.
  Lett.} {\bfseries 123} (Dec, 2019) 250501}.
  \url{https://link.aps.org/doi/10.1103/PhysRevLett.123.250501}.

\bibitem{arunachalam2015robustness}
S.~Arunachalam, V.~Gheorghiu, T.~Jochym-O’Connor, M.~Mosca, and P.~V.
  Srinivasan, ``On the robustness of bucket brigade quantum ram,'' {\em New
  Journal of Physics} {\bfseries 17} no.~12, (2015) 123010.

\bibitem{liu2022quantum}
J.~Liu, C.~T. Hann, and L.~Jiang, ``Quantum data center: Theories and
  applications,'' {\em arXiv preprint arXiv:2207.14336} (2022) .

\bibitem{lapworth2022hybrid}
L.~Lapworth, ``A hybrid quantum-classical cfd methodology with benchmark hhl
  solutions,'' {\em arXiv preprint arXiv:2206.00419} (2022) .

\bibitem{qiskit}
\url{https://qiskit.org}.

\bibitem{childs2017quantum}
A.~M. Childs, R.~Kothari, and R.~D. Somma, ``Quantum algorithm for systems of
  linear equations with exponentially improved dependence on precision,'' {\em
  SIAM Journal on Computing} {\bfseries 46} no.~6, (2017) 1920--1950.

\bibitem{woerner2019quantum}
S.~Woerner and D.~J. Egger, ``Quantum risk analysis,'' {\em npj Quantum
  Information} {\bfseries 5} no.~1, (2019) 1--8.

\bibitem{vazquez2022enhancing}
A.~C. Vazquez, R.~Hiptmair, and S.~Woerner, ``Enhancing the quantum linear
  systems algorithm using richardson extrapolation,'' {\em ACM Transactions on
  Quantum Computing} {\bfseries 3} no.~1, (2022) 1--37.

\bibitem{qbraid}
\url{https://qbraid.com}.

\end{thebibliography}\endgroup
\end{document}